\documentclass[pdflatex,sn-basic,iicol]{sn-jnl}
\usepackage{graphicx}%
\usepackage{multirow}%
\usepackage{amsmath,amssymb,amsfonts}%
\usepackage{amsthm}%
\usepackage{mathrsfs}%
\usepackage[title]{appendix}%
\usepackage{xcolor}%
\usepackage{textcomp}%
\usepackage{manyfoot}%
\usepackage{booktabs}%
\usepackage{algorithm}%
\usepackage{algorithmicx}%
\usepackage{algpseudocode}%
\usepackage{listings}%
\usepackage{csquotes}
\usepackage{placeins}
\usepackage{marvosym}
\usepackage{graphicx}
\usepackage{caption}
\usepackage{array} 
\usepackage{longtable}
\usepackage{float}
\usepackage{natbib}

\newcommand{\rr}{\color{green}$\checkmark$}

\newcommand{\ww}{\color{red}$\times$}

\begin{document}
\title[Article Title]{ChatHTTPFuzz: Large Language Model-Assisted IoT HTTP Fuzzing}

\author[1]{\fnm{Zhe} \sur{Yang}} 

\author*[1]{\fnm{Hao} \sur{Peng}}\email{penghao@buaa.edu.cn}

\author[1]{\fnm{Yanling} \sur{Jiang}} 

\author[2]{\fnm{Xingwei} \sur{Li}} 

\author[1]{\fnm{Haohua} \sur{Du}} 

\author[3]{\fnm{Shuhai} \sur{Wang}} 

\author[1]{\fnm{Jianwei} \sur{Liu}}

\affil[1]{\orgdiv{School of Cyber Science and Technology}, \orgname{Beihang University}, \orgaddress{\city{Beijing}, \postcode{100191}, \country{China}}}

\affil[2]{\orgname{Information Engineering University}, \orgaddress{\city{Zhengzhou}, \postcode{10587}, \state{Henan}, \country{China}}}

\affil[3]{\orgname{College of Information Science and Technology}, \orgname{Shijiazhuang Tiedao University}, \orgaddress{\city{Shijiazhuang}, \postcode{050043}, \state{Hebei}, \country{China}}}

\abstract{
Internet of Things (IoT) devices offer convenience through web interfaces, web VPNs, and other web-based services, all relying on the HTTP protocol.
However, these externally exposed HTTP services present significant security risks.
Although fuzzing has shown some effectiveness in identifying vulnerabilities in IoT HTTP services, most state-of-the-art tools still rely on random mutation strategies, leading to difficulties in accurately understanding the HTTP protocol's structure and generating many invalid test cases.
Furthermore, These fuzzers rely on a limited set of initial seeds for testing. 
While this approach initiates testing, the limited number and diversity of seeds hinder comprehensive coverage of complex scenarios in IoT HTTP services.
In this paper, we investigate and find that large language models (LLMs) excel in parsing HTTP protocol data and analyzing code logic. 
Based on these findings, we propose a novel LLM-guided IoT HTTP fuzzing method, ChatHTTPFuzz, which automatically parses protocol fields and analyzes service code logic to generate protocol-compliant test cases. 
Specifically, we use LLMs to label fields in HTTP protocol data, creating seed templates. 
Second, The LLM analyzes service code to guide the generation of additional packets aligned with the code logic, enriching the seed templates and their field values. 
Finally, we design an enhanced Thompson sampling algorithm based on the exploration balance factor and mutation potential factor to schedule seed templates.
We evaluate ChatHTTPFuzz on 14 different real-world IoT devices. 
It finds more vulnerabilities than SNIPUZZ, BOOFUZZ, and MUTINY. 
ChatHTTPFuzz has discovered 103 vulnerabilities, of which 68 are unique, and 23 have been assigned CVEs.
}

\maketitle

\section{Introduction}\label{sec:intro}
Internet of things (IoT) technology is rapidly advancing, transforming people's lives and work environments in unprecedented ways \citep{ezechina_internet_nodate}.
As of 2024, there are 17.02 billion IoT-connected devices worldwide, expected to reach approximately 30 billion by 2030 \citep{IoT2024}.
However, this exponential growth in IoT devices presents significant security challenges. 
To facilitate user management and ease of use, these interconnected devices offer various interaction methods, the most common of which is the web interface \citep{kumar_unified_2016}.
While web services provide convenience, they also introduce significant security risks. 
Many IoT devices have design flaws in this area, harboring numerous security vulnerabilities \citep{pourrahmani_review_2023,lindsey_odonnell_more_2020}.
Exploiting vulnerabilities in these interfaces can lead to unauthorized access, data theft, or remote code execution (RCE) attacks \citep{noman_code_2023,siwakoti_advances_2023}. 
There are numerous research methods for the security analysis of these IoT devices. 
Commonly used vulnerability detection techniques include static analysis, dynamic analysis, binary comparison, and fuzzing \citep{cui_empirical_2022}.
As software scale and complexity increase, fuzzing exhibits unparalleled advantages over other vulnerability detection techniques \citep{liang_fuzzing_2018}.

Fuzzing, as an effective vulnerability detection technique, plays a crucial role in the security testing of IoT devices. 
However, due to the complexity of the HTTP protocol and the unique hardware and software characteristics of IoT devices, existing fuzzing tools and methodologies still face numerous limitations and challenges in practical applications.
The strict protocol structure of HTTP requires fuzzing tools to generate test data that conforms to protocol specifications. 
However, many current fuzzing tools face inefficiencies when handling HTTP protocol packets. 
Tools like FirmAFL \citep{zheng_firm-afl_2019} and Mutiny \citep{mutiny_2017} use byte-level mutation methods, randomly mutating the bytes of HTTP packets. 
Although this approach is simple, it generates many invalid seeds that do not conform to protocol standards, failing to elicit valid responses from the server and thus reducing fuzzing efficiency. 
Tools like BooFuzz \citep{jtpereyda_jtpereydaboofuzz_nodate} and Peach \citep{peach} utilize structured seed templates to regulate the mutation process, ensuring the generated data adheres more closely to the requirements of the HTTP protocol. 
However, these tools heavily rely on manually created seed templates, requiring the manual definition of mutation fields and types.
This experience-dependent approach is time-consuming and labor-intensive, significantly limiting automation and applicability when dealing with complex or unfamiliar protocols.
Additionally, while SNIPUZZ \citep{feng_snipuzz_2021} attempts to infer the message structure of test cases by combining response information to improve the efficiency of byte-level mutations, this method remains inefficient regarding protocol structure awareness.

Advanced black-box and grey-box fuzzers for IoT devices focus primarily on leveraging limited feedback information to guide the fuzzing process \citep{zheng_firm-afl_2019, feng_snipuzz_2021} 
while paying insufficient attention to the acquisition and quality of initial seeds. 
For instance, tools like TriforceAFL \citep{ncc-group_triforce_2017} and Firm-AFL concentrate on simulating or re-hosting IoT firmware to gather more testing information, enhancing the efficiency of fuzzing and the ability to detect vulnerabilities. 
Labrador \citep{liu_labrador_nodate} collects feedback information, such as code coverage and distance, under black-box fuzzing of IoT devices to guide the testing process. 
However, if the initial seed set is insufficient or lacks representativeness, comprehensive vulnerability detection remains challenging even with efficient feedback mechanisms and optimization strategies. 
Therefore, acquiring and optimizing initial seeds requires further research in advanced fuzzing methodologies.

In this paper, our investigation into the capabilities of large language models (LLMs) \citep{LLM_survey} reveals that LLMs achieve a 0\% error rate and a false negative rate of only 4.74\% when parsing HTTP protocol fields. 
Additionally, they achieve 98.58\% coverage of packet fields when analyzing code logic.
These results indicate that LLMs possess excellent automated parsing capabilities, enabling protocol-aware field annotation and backend code analysis without manual intervention. 
Therefore, we propose an optimized HTTP fuzzing method for IoT devices based on Large Language Models (LLMs). 
First, we leverage the protocol parsing capabilities of LLMs to identify suitable mutation points within HTTP protocol data. 
Second, we utilize LLMs' code analysis and protocol generation abilities to generate data packets aligned with the code logic, thereby enhancing both the quantity and quality of seeds. 
Finally, we design an enhanced Thompson sampling algorithm, incorporating a dual-factor gain (the exploration balance factor and mutation potential factor), to efficiently schedule seed templates.
Specifically, ChatHTTPFuzz employs three core technologies to address IoT HTTP protocol fuzzing challenges.

\textbf{LLM-Guided HTTP protocol variable annotation (LVA) Technique} is employed to overcome the limitations in protocol structure analysis. 
This method employs preset prompt instructions to drive a Large Language Model (LLM) to annotate variables within HTTP protocol data accurately. 
By introducing the concept of seed templates, we enhance seed quality and increase mutation legality while defining mutation data types based on variable type attributes. 
In the experiments of Section \ref{subsec3.1}, we test 140 constructed parameter-passing data packets. 
The results demonstrate that the LLM achieves a zero false-negative rate in recognizing various parameter formats, including XML, JSON, and key-value pairs.
This technique effectively guides annotation tasks in fuzzing, significantly improving testing accuracy and efficiency.

\textbf{LLM-Guided Seed Template Enrichment (LSTE) Algorithm} is employed to address the issue of limited sources for obtaining initial seeds.
This algorithm improves the seed template library using two methods: packet expansion and field value enrichment. 
The packet expansion method deals with situations with no corresponding route codes by using an LLM to interpret statically analyzed backend code and generate new packets. 
The field value enrichment method extracts potential value sets for variable fields by analyzing the field processing logic in the backend code based on existing packets and seed templates. 
These two methods allow ChatHTTPFuzz to create more effective seed templates based on code semantics, significantly improving seed quality. 
The LSTE algorithm not only expands testing coverage but also enhances the efficiency and accuracy of fuzzing, providing a more reliable foundation for discovering potential security vulnerabilities.

\textbf{Dual-Factor Gain Seed Template Scheduling Algorithm (STSA) based on reinforcement learning} is employed to tackle the challenges of seed scheduling.
This algorithm enhances traditional Thompson Sampling \citep{karamcheti_adaptive_2018} by introducing a Mutation Potential Factor ($M_{factor}$) and an Exploration Balance Factor ($E_{factor}$), thereby incorporating heuristic information. 
The $M_{factor}$ evaluates the potential mutation capability of templates, enabling the algorithm to consider the value of templates in future iterations proactively. 
The $E_{factor}$, derived from the usage history of templates, ensures an effective balance between exploration and exploitation during decision-making. 
The integration of these two factors accounts for historical performance and balances the evaluation of future potential and global trade-offs, resulting in a more efficient and balanced selection strategy.

We have implemented the prototype system ChatHTTPFuzz and conducted a comprehensive evaluation on 14 models of IoT devices. 
The experimental results demonstrate that ChatHTTPFuzz significantly outperforms three advanced vulnerability detection tools: Boofuzz \citep{jtpereyda_jtpereydaboofuzz_nodate}, a template-based mutation fuzzer; Snipuzz \citep{feng_snipuzz_2021}, a black-box fuzzer for IoT; and Mutiny \citep{mutiny_2017}, an efficient network protocol fuzzer. 
Comparative experiments show that ChatHTTPFuzz successfully identifies 103 security vulnerabilities across 14 IoT devices, 13 times the number of vulnerabilities detected by Boofuzz, Snipuzz, and Mutiny. Among these vulnerabilities, 68 are previously undisclosed, and 23 receive official CVE identifiers.
In summary, this paper makes the following contributions.

$\bullet$
We propose a novel HTTP protocol data awareness method based on LLMs for the automated annotation of protocol variable data, seed template generation, and guidance for seed mutation. Accordingly, we also introduce an algorithm for enriching the seed template library to improve the quality and quantity of seeds.

$\bullet$
We design an innovative dual-factor gain-enhanced Thompson sampling algorithm (based on an exploration balancing factor and mutation potential factor) to schedule seed templates using information from the seed templates.

$\bullet$
We evaluate ChatHTTPFuzz on 14 different real-world IoT devices. It identifies more vulnerabilities than the baseline methods, discovering 103 vulnerabilities, 59 of which are unique, with 23 assigned CVEs.

The structure of this work is as follows.
An in-depth exploration of the HTTP protocol and fuzzing techniques, along with an analysis of the potential applications and advantages of large language models (LLMs) in fuzzing, is provided in Section \ref{sec:backg}.
The evaluation of the LLM’s ability to parse and generate HTTP protocols through experiments assessing its performance in processing HTTP data and interpreting server-side code is presented in Section \ref{sec:invest}.
Drawing on the results of these experiments, we design the ChatHTTPFuzz system, leveraging LLMs for dynamic HTTP fuzzing with enhanced capabilities in data parsing, seed generation, and scheduling optimization, in Section \ref{sec:design}.
A detailed discussion of the key implementation modules, focusing on the system’s core functionalities and architectural design, in Section \ref{sec:impl}.
Finally, a comprehensive performance evaluation, comparing the system with state-of-the-art (SOTA) solutions across multiple metrics (efficiency, and vulnerability detection) in Section \ref{sec6}.
\section{BackGround}\label{sec:backg}

We begin by introducing the techniques of HTTP protocol fuzzing in the context of IoT environments, along with the key challenges associated with this approach. 
Following this, we explore the technical capabilities and motivations identified within large language models (LLMs) for addressing these challenges.

\subsection{HTTP Protocol Fuzzing}\label{subsec2.1}
In the design and implementation of IoT devices, the open-source HTTP protocol is often employed as the web service interface due to its simplicity and compatibility. 
According to RFC 2616  \citep{noauthor_hypertext_nodate}, the HTTP protocol is strictly defined across a 176-page document. 
Fuzzing the HTTP protocol is a technique used to discover vulnerabilities and security issues by sending large amounts of random or malformed data to a target application \citep{kallus_http_2024}. 
HTTP protocol fuzzing requires fuzzers to comply with protocol specifications when generating test data strictly. 
Mutating fixed fields of the protocol can result in many invalid packets that cannot be parsed, leading to low fuzzing efficiency \citep{tsai_rest_2021}. 
Additionally, if the number of initial seeds is insufficient or does not match the expected fields of the target code, it becomes difficult to test more functionalities and code branches within the service.

\subsection{Large Language Models}
\label{subsec2.2}

Large language models (LLMs) are deep learning-based models trained using massive amounts of data and powerful computational resources. 
They can capture complex language patterns and structures and perform exceptionally well in natural language processing tasks and text generation \citep{brown_language_2020}.
In recent years, the potential of large language models (LLMs) has extended beyond traditional language processing, increasingly emerging in areas such as protocol parsing and code analysis \citep{fan_automated_2023, jain_jigsaw_2022}.
Network protocols are typically defined by standard documents such as RFCs (Request for Comments), which use natural language to describe various details of the protocols. 
Since LLMs are trained on vast amounts of natural language data, they can understand the language and content of these documents and generate relevant protocol content based on the standards or parsing complex structures within protocols. 
Moreover, LLMs can comprehend code intent through natural language prompts and automatically generate code snippets \citep{du_evaluating_2024}.
LLMs' high degree of automation and intelligence enables them to handle complex tasks efficiently. 
LLMs can understand and execute various tasks by utilizing natural language prompts. 
These capabilities can potentially address open challenges in IoT HTTP fuzzing.

\subsection{LLMs in Fuzzing}
\label{subsec2.3}

Some existing work has already applied LLMs to fuzzing. 
For example, GPTFuzzer uses human-written templates as initial input and leverages LLMs to generate new templates through mutation. 
ChatAFL \citep{meng_large_2024} utilizes LLMs to parse the RTSP protocol, mark variable fields, guide the mutation process, and predict protocol sequences. 
mGPTFuzz \citep{ma_one_2024} relies on LLMs to convert over a thousand pages of specification documents into machine-readable formats, performing protocol state analysis and generation.
Fuzz4All \citep{xia_fuzz4all_2024} uses LLMs' code analysis capabilities to enhance fuzzing capabilities for \enquote{systems under test} (such as compilers, runtime engines, constraint solvers, and software libraries). 
These models have helped accomplish many complex tasks traditionally requiring human expertise and have demonstrated significant results in related research, offering new approaches to addressing challenges in IoT HTTP fuzzing. 
In this paper, we leverage the Prompt Tuning method \citep{chen_unleashing_2024} to utilize LLMs for guiding HTTP protocol fuzzing. 

\section{Capability Evaluation: The Impact of Large Language Models on HTTP Protocol Fuzzing}
\label{sec:invest}

This study conducts two experiments to assess the ability of Large Language Models in parsing and generating HTTP protocol.
The first experiment was based on the RFC HTTP 1.1 protocol specification, constructing standard-compliant HTTP packets and using the GPT-4o API \citep{gpt4o} with preset prompts for parsing and generation. 
Experts manually reviewed the generated results to assess their compliance with the protocol and overall validity. 
The second experiment focused on backend code analysis, where the GPT-4o API was used to perform an in-depth analysis of the target system’s code, generating new HTTP packets based on this analysis. 
These generated packets were then sent to the target web service to verify the effectiveness of parameter parsing.

The above experiments aim to systematically evaluate the accuracy and adaptability of large language models in HTTP protocol parsing and generation. 
The specific prompt sequences and the method of guiding fuzzing are discussed in Sections \ref{subsec4.1} and  \ref{subsec4.2}.


\subsection{Field Identification: Comprehensive and Precise Annotation}
\label{subsec3.1}
This experiment systematically considered the HTTP protocol's various parameter transmission methods and types.
By thoroughly studying the RFC 2616 specification, we carefully design and construct 140 representative parameter transmission packets. 
To ensure accuracy and diversity, the research team dedicates substantial time and effort to verify the correctness of each transmission method and parameter type. 

\begin{table}[!tb]
    \centering
    \caption{Types and Subtypes of Parameters in Data Packets. Includes 5 major categories and 14 subcategories.}
    \label{tab:parameter_data_packets}
    \small
    \begin{tabular}{lc} 
        \toprule
        \textbf{Type} & \textbf{Description} \\
        \midrule
        Path & URL Path Params \\
        \midrule
        \multirow{3}{*}{Query String} 
            & Key-Value Pair Params \\
            & JSON Params \\
            & XML Params \\
        \midrule
        \multirow{8}{*}{Request Body} 
            & Form Data \\
            & JSON format(Form Data) \\
            & XML format(Form Data) \\
            & Multipart Form Data \\
            & JSON format(Multipart) \\
            & XML format(Multipart) \\
            & JSON format \\
            & XML format \\
        \midrule
        Header & Header Params \\
        \midrule
        Cookie & Cookie Params \\
        \bottomrule
    \end{tabular}
    
\end{table}

The packets are categorized into five main groups based on the method of transmission: path parameters, query string parameters, request body parameters, header parameters, and cookie parameters. 
Furthermore, these packets are systematically divided into 14 distinct subcategories based on the variation in parameter types, as illustrated in Table \ref{tab:parameter_data_packets}.
We cover the most commonly encountered transmission methods among the 140 parameter packets constructed across 14 types. 
In addition to standard data structures, we also introduce several non-standard custom data structures to ensure the comprehensiveness and diversity of the testing process. 
The data types are detailed in Table \ref{tab:examples_data_structures}.
Among them, types 1, 2, and 3 represent standard data structures, types 5, 6, and 7 are custom data structures, and types 8, 9, and 10 are composite data structures.

\begin{table*}[!tb]
    \centering
    \caption{Examples of Different Data Structures. Including key-value pairs, JSON, XML, custom formats, and composite data.}
    \label{tab:examples_data_structures}
    \small
    \begin{tabular}{lcl}
        \toprule
        \textbf{ID} & \textbf{Type} & \textbf{Example} \\
        \toprule
        1  & Key-Value & name=alice\&age=10 \\
        2  & JSON & \{\enquote{name}: \enquote{alice}, \enquote{age}: 10\} \\
        3  & XML & \textless{}user\textgreater{}\textless{}name\textgreater{}alice\textless{}/name\textgreater{}\textless{}/user\textgreater{} \\
        4  & Array & Array=[\enquote{alice},10] \\
        5  & Custom 1 & data=alice:10 \\
        6  & Custom 2 & data=alice|10 \\
        7  & Custom 3 & data=alice-10 \\
        8  & Other 1 & Data=\{\enquote{info}: [\enquote{alice}, \enquote{alice:10}]\} \\
        9  & Other 2 & Data=\textless{}user\textgreater{}\textless{}info\textgreater{}alice|stu\textless{}/info\textgreater{}\textless{}/user\textgreater{} \\
        10 & Other 3 & Data=[\enquote{stu},\enquote{alice:10},\enquote{bob-20}] \\
        \bottomrule
    \end{tabular}
\end{table*}

This study systematically analyzes 140 carefully designed parameter transmission packets through experimental evaluation.
Using a quantitative analysis approach, we record and compare three key metrics: the original number of parameters per packet, the number of parameters successfully parsed by the LLM, and the number of misidentified parameters. 
We further calculate the false negative and false positive rates based on this data. 
This multi-dimensional analysis provides a basis for assessing the performance of the GPT-4o API in HTTP protocol parsing. 
\begin{table*}[!tb]
    \centering
    \caption{Parameter Recognition and Error Rates. NP: Number of Parameters, CIC: Correctly Identified Count, IIC: Incorrectly Identified Count, FNR: False Negative Rate, FPR: False Positive Rate.}
    \label{tab:parameter_recognition}
    \small
    \begin{tabular}{lccccc} 
        \toprule 
        \textbf{Type} & \textbf{NP} & \textbf{CIC} & \textbf{IIC} & \textbf{FNR} & \textbf{FPR} \\
        \midrule 
        URL Path Parameters & 44 & 44 & 0 & 0\% & 0\% \\
        Key-Value Pair Parameters & 43 & 43 & 0 & 0\% & 0\% \\
        JSON Parameters & 41 & 41 & 0 & 0\% & 0\% \\
        XML Parameters & 43 & 43 & 0 & 0\% & 0\% \\
        Form Data & 45 & 45 & 0 & 0\% & 0\% \\
        Form Data (JSON) & 45 & 45 & 0 & 0\% & 0\% \\
        Form Data (XML) & 41 & 41 & 0 & 0\% & 0\% \\
        Multipart Form Data & 55 & 55 & 10 & 0\% & 15.38\% \\
        Multipart Form Data (JSON) & 43 & 43 & 10 & 0\% & 18.87\% \\
        Multipart Form Data (XML) & 55 & 55 & 10 & 0\% & 15.38\% \\
        JSON format & 42 & 42 & 0 & 0\% & 0\% \\
        XML format & 41 & 41 & 0 & 0\% & 0\% \\
        Header Parameters & 30 & 30 & 0 & 0\% & 0\% \\
        Cookie Parameters & 35 & 35 & 0 & 0\% & 0\% \\
        \midrule 
        \textbf{Total} & 603 & 603 & 30 & 0\% & 4.74\% \\
        \bottomrule 
    \end{tabular}
\end{table*}

Experimental data from Table \ref{tab:parameter_recognition} demonstrate that the GPT-4 API performs exceptionally well in HTTP parameter parsing. 
Notably, it achieved perfect accuracy, with both false negative and false positive rates of 0\%, when handling URL path parameters, key-value pair parameters, and JSON parameters. 
However, when parsing complex multipart form data, the API exhibited certain limitations, with false positive rates of 15.38\%, 18.87\%, and 15.38\%, respectively. 
This indicates that the API may mistakenly identify non-parameter data as variables when dealing with highly structured form data.

Importantly, the GPT-4o API excelled in processing custom-formatted parameters, highlighting its potential to adapt to non-standard data structures. 
Despite room for improvement in specific complex scenarios, the GPT-4o API demonstrates outstanding overall performance and adaptability in HTTP parameter recognition and parsing.

\subsection{Package Enrichment: Parsing and Generation}

One of the core strengths of large language models (LLMs) lies in their exceptional code-handling capabilities, which stem from extensive learning of vast amounts of code during training \citep{tufano_towards_2021}. 
To apply LLMs' code parsing and generation abilities to HTTP protocol fuzzing, this study proposes two methods for expanding the existing seed corpus. 

\begin{figure*}[!ht]
\centering
\includegraphics[width=0.8\linewidth]{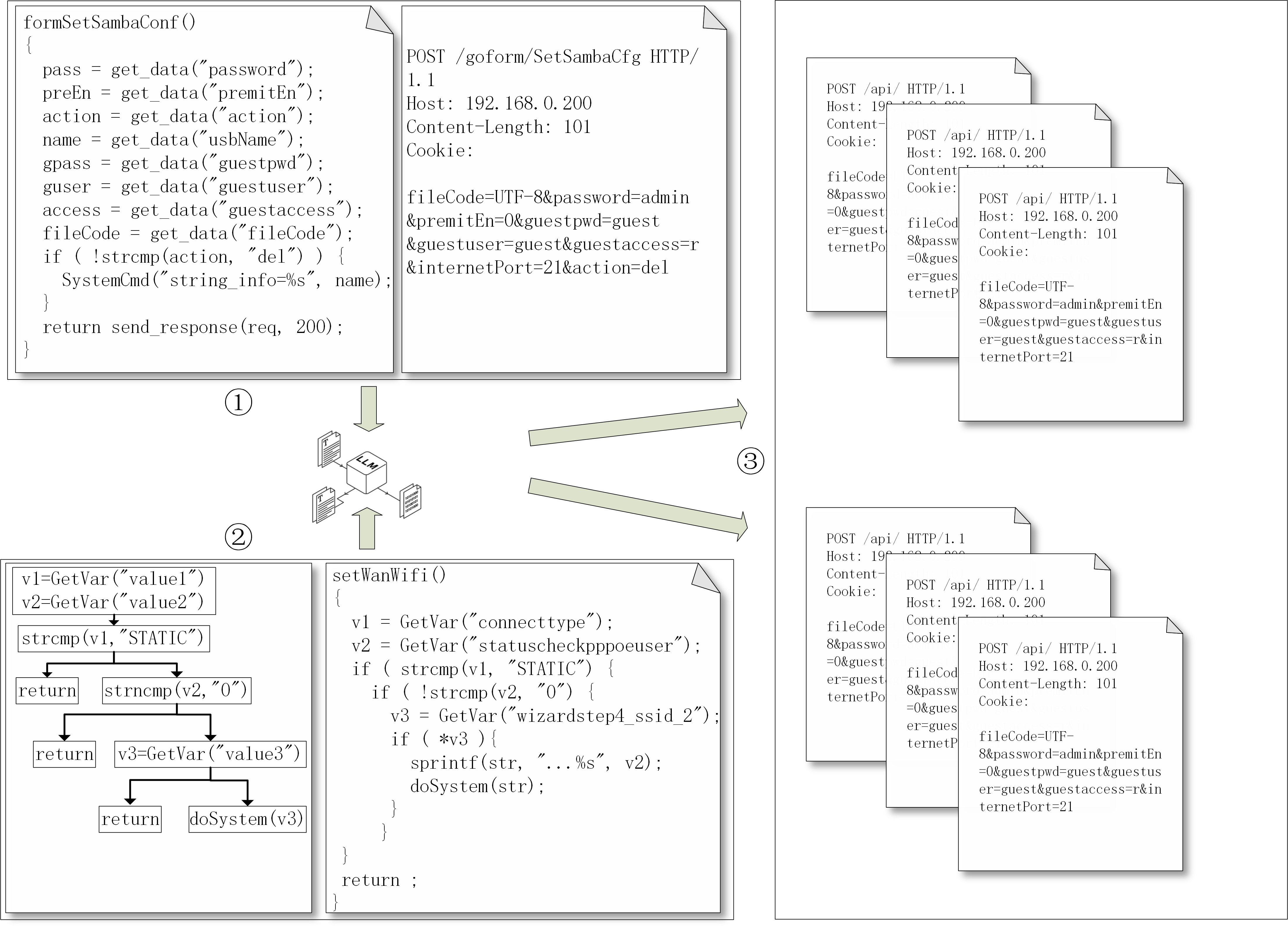}
\caption{LLM-Assisted Packet Generation. The Upper Part of the Diagram Contains Packets and Code, While the Lower Part Contains Code Only. The Code in the Green Section is Key to Determining the Branching Direction.}\label{Code Analysis}
\end{figure*}

The first method relies on a given code snippet and its corresponding data packet, where the LLM analyzes the code's logic and field names to generate an extended version of the data packet (referred to as Type 0). 
Without a corresponding data packet, the second method uses the provided code snippet to allow the LLM to generate entirely new protocol data packets that align with the code's logic (referred to as Type 1).

As shown in Figure \ref{Code Analysis}, Section 1 demonstrates how the LLM, in the presence of a given data packet, generates new packets by analyzing the backend code.
In traditional fuzzing techniques, relying solely on data mutation cannot set the action field in a packet to “del” for two main reasons: first, the original packet does not contain this field, and second, fuzzing itself cannot anticipate that the value “del” might trigger a new code branch.
However, by parsing the backend code, the LLM intelligently identifies critical input parameters and predicts how different parameter values could influence the backend logic. 
Consequently, the LLM can generate targeted data packets, such as those with $action=del$, to trigger the delete operation's code branch. 
The second part demonstrates that, with only the code available, the LLM generates a set of data packets consistent with the code logic by analyzing field parsing and code branch structures.
LLMs can identify deep logical branches that can only be reached through multiple consecutive conditional evaluations.
In such cases, the LLM generates data packets that satisfy single conditions and constructs a sequence of input values to trigger code execution paths hidden behind complex logical structures. 
For instance, when $v1=STATIC$ and $v2=0$ in this code, further parsing of wizardstep4 is required to trigger the command execution.

We design the following experiment for three Internet of Things (IoT) devices in response to the two scenarios above. 
We select ten service codes for each device with existing data packets and ten without data packets. 
To assess the validity of the generated data packets, we sent the generated data to the corresponding devices and utilized dynamic debugging \citep{noauthor_gdb_nodate} to analyze whether the variable values in the sent data packets were assigned to the corresponding parameters at the code level. 
This approach allowed us to determine the effectiveness of the data packet generation. 
Based on the evaluation of 60 data packets across three devices, the experimental results in Table \ref{tab:field_recognition} show that the average field recognition rate of these devices reached 98.58\%.

\begin{table}[!tb]
    \centering
    \caption{Field Recognition Rates for Different IoT Devices. DN: Device Name, GT: Generation Type, PC: Packet Count, FC: Field Count, RC: Recognized Count, RR: Recognition Rate.}
    \label{tab:field_recognition}
    \begin{tabular}{lccccc}
        \toprule
        \textbf{DN} & \textbf{GT} & \textbf{PC} & \textbf{FC} & \textbf{RC} & \textbf{RR} \\
        \midrule
        Cisco RV110W & 0 & 10 & 34 & 33 & 97.06\% \\
                     & 1 & 10 & 22 & 22 & 100\% \\
        Tenda AC15   & 0 & 10 & 16 & 16 & 100\% \\
                     & 1 & 10 & 23 & 22 & 95.65\% \\
        Dlink Dir-816 & 0 & 10 & 20 & 20 & 100\% \\
                      & 1 & 10 & 26 & 26 & 100\% \\
        \midrule
        \textbf{Total} & & 60 & 141 & 139 & 98.58\% \\
        \bottomrule
    \end{tabular}
\end{table}

In conclusion, this study systematically evaluates the capabilities of large language models (LLMs) in HTTP protocol parsing, code analysis, and protocol generation. 
By testing 140 data packets encompassing various parameter passing methods and parameter types, the results demonstrate that LLMs achieve high accuracy in recognizing and parsing common and custom data structures. 
Additionally, the experiments confirmed the potential of LLMs in code analysis and protocol data generation. 
By generating data packets that align with the given code logic, LLMs can effectively expand existing seed libraries, enhancing the efficiency of fuzzing and increasing code coverage. 
These findings indicate that LLMs are promising for network protocol testing and code analysis applications.
Building on these strengths, our next step is to design and implement a system named ChatHTTPFuzz, leveraging LLMs to further improve the intelligence and automation of IoT HTTP protocol fuzzing.
\section{Design of ChatHTTPFuzz}
\label{sec:design}
\begin{figure*}[!ht]
\centering
\resizebox{\textwidth}{!}{
  \includegraphics{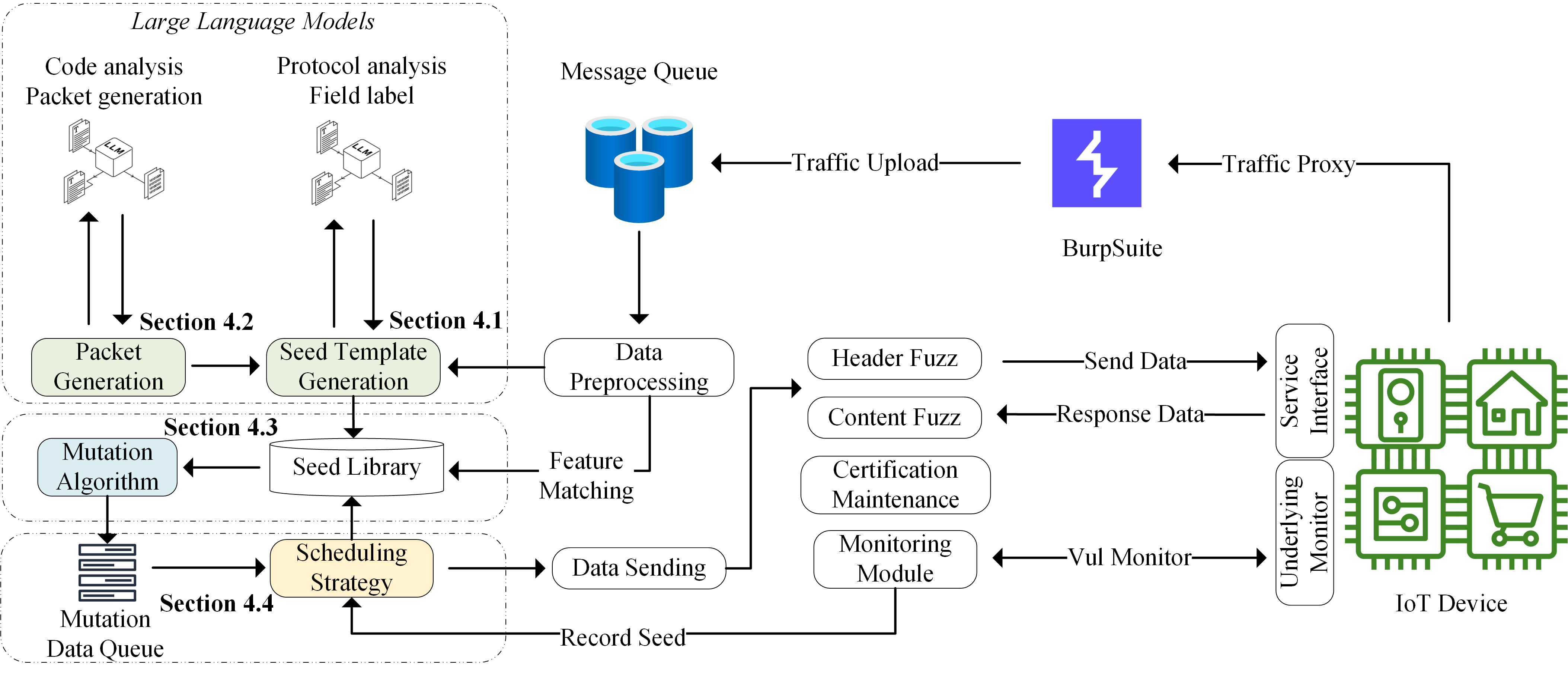}
}
\caption{ChatHTTPFuzz workflow. 
The green section in the diagram represents our innovation, utilizing LLM for protocol parsing, seed template generation, and packet enrichment.
The diagram also includes modules for seed mutation and scheduling, with the system's operational infrastructure shown on the right.}\label{ChatHTTPFuzz}
\end{figure*}

The LLM demonstrated exceptional protocol parsing and generation capabilities in the abovementioned experiments. 
To tackle the challenges of HTTP service fuzzing in the Internet of Things (IoT) environment, this paper presents a new approach for fuzzing HTTP protocols in IoT devices, known as ChatHTTPFuzz, guided by LLM technology.
ChatHTTPFuzz effectively addresses key issues such as field identification and seed generation by leveraging advanced LLM capabilities. 
Figure \ref{ChatHTTPFuzz} provides a detailed illustration of the ChatHTTPFuzz workflow.

To apply large language models (LLMs) in HTTP protocol fuzzing, we first propose a method for leveraging LLMs to understand the structure of the HTTP protocol (Section \ref{subsec4.1}). 
We also design a sequence of specific prompts for HTTP protocol parsing (Section \ref{subsec4.1.1}) to improve the accuracy of LLMs in parsing and annotation tasks. 
Additionally, we develop a seed template structure that stores key information and guides the mutation process (Section \ref{subsec4.1.2}). 
Leveraging LLM's code parsing and protocol generation capabilities, combined with static code analysis and relevant data packets, we generate more initial seeds that align with the backend code logic (Section \ref{subsec4.2}). 
Based on field types and mutation spaces (field value sets), ChatHTTPFuzz mutates the seed templates to generate test seeds (Section \ref{subsec4.3}). 
Finally, by analyzing the limited response content from the server, we design an enhanced Thompson sampling seed template scheduling algorithm based on dual-factor gain to balance the exploration and exploitation between seed templates (Section \ref{subsec4.4}). 

\subsection{LLM-Guided HTTP Protocol Structure Aware}
\label{subsec4.1}

In this section, we will use the Prompt Tuning method to leverage a large language model (LLM) to annotate fuzzing variables in request data and understand the protocol data structure, guiding the mutation process more effectively.

\subsubsection{LLM-Guided HTTP Protocol Variable Annotation}
\label{subsec4.1.1}

A complete HTTP protocol consists of two main components: the request header and the request body. 
While the request header does not involve complex data structures, it contains various fields, making it challenging to identify and process using existing rules accurately. 
As the core of data transmission, the request body typically includes complex data structures and custom-formatted data, increasing the difficulty of recognition and parsing. 
To enhance the accuracy of variable identification within the protocol using LLMs, we employ the following methods to augment the capabilities of the LLM.

In large language models (LLMs), the \enquote{temperature} parameter regulates the randomness of generated text. 
Higher temperatures produce more creative and diverse outputs, while lower temperatures produce more precise and stable results. 
To ensure the accurate extraction of information from events, we typically set the temperature to 0 \citep{renze_effect_2024}.

\begin{figure*}[!ht]
\centering
\includegraphics[width=0.8\linewidth]{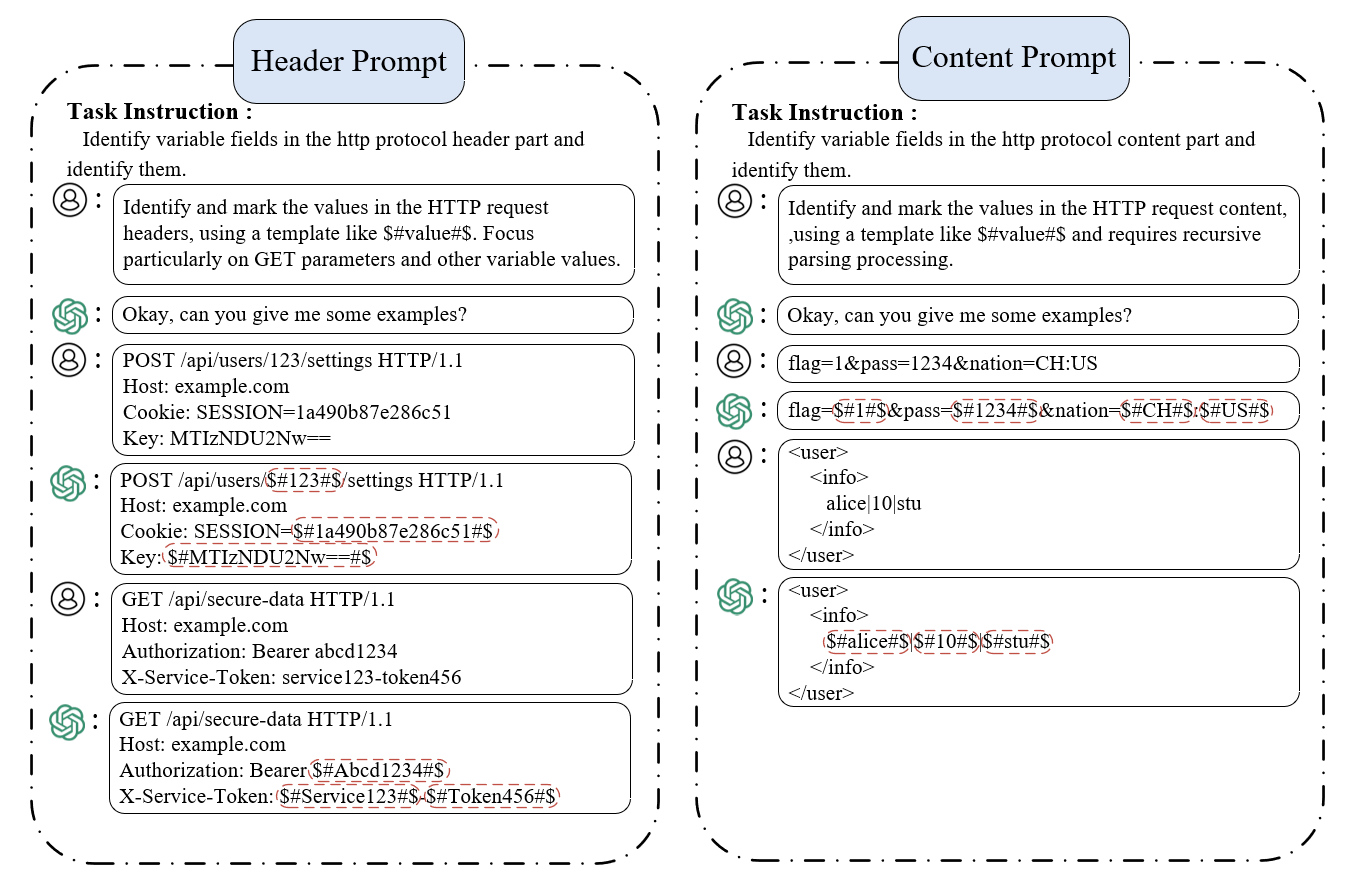}
\caption{Package Prompt. On the left are prompts for annotating Header variables, while on the right are prompts designed for Content.}\label{Package Prompt}
\end{figure*}

To enhance task specificity and accuracy, we divide the complete HTTP packet into two parts: the request header (Header) and the request body (Content). 
Different prompts are designed to address the unique characteristics of each part, guiding the LLM in understanding the syntactic structure and producing the desired output. 
Figure \ref{Package Prompt} illustrates the two prompts for labeling Header and Content. 
The Header prompt directs the model’s attention to variable fields beyond the protocol format, while the Content prompt focuses on the values of variables within the data structure.

To further ensure accuracy and consistency, we adopt a few-shot learning approach, using multiple question-and-answer examples rather than relying on a single instance to clarify the task requirements. 
The examples specify which fields are eligible for labeling and which are essential components of the protocol and data structure that must remain unchanged. 
In both the prompts and the examples, the $\$\#\#\$$ symbol marks variable elements, serving as a data source for generating seed templates.

\subsubsection{Seed Template Structure}
\label{subsec4.1.2}
\begin{figure*}[!ht]
\centering
\includegraphics[width=0.8\linewidth]{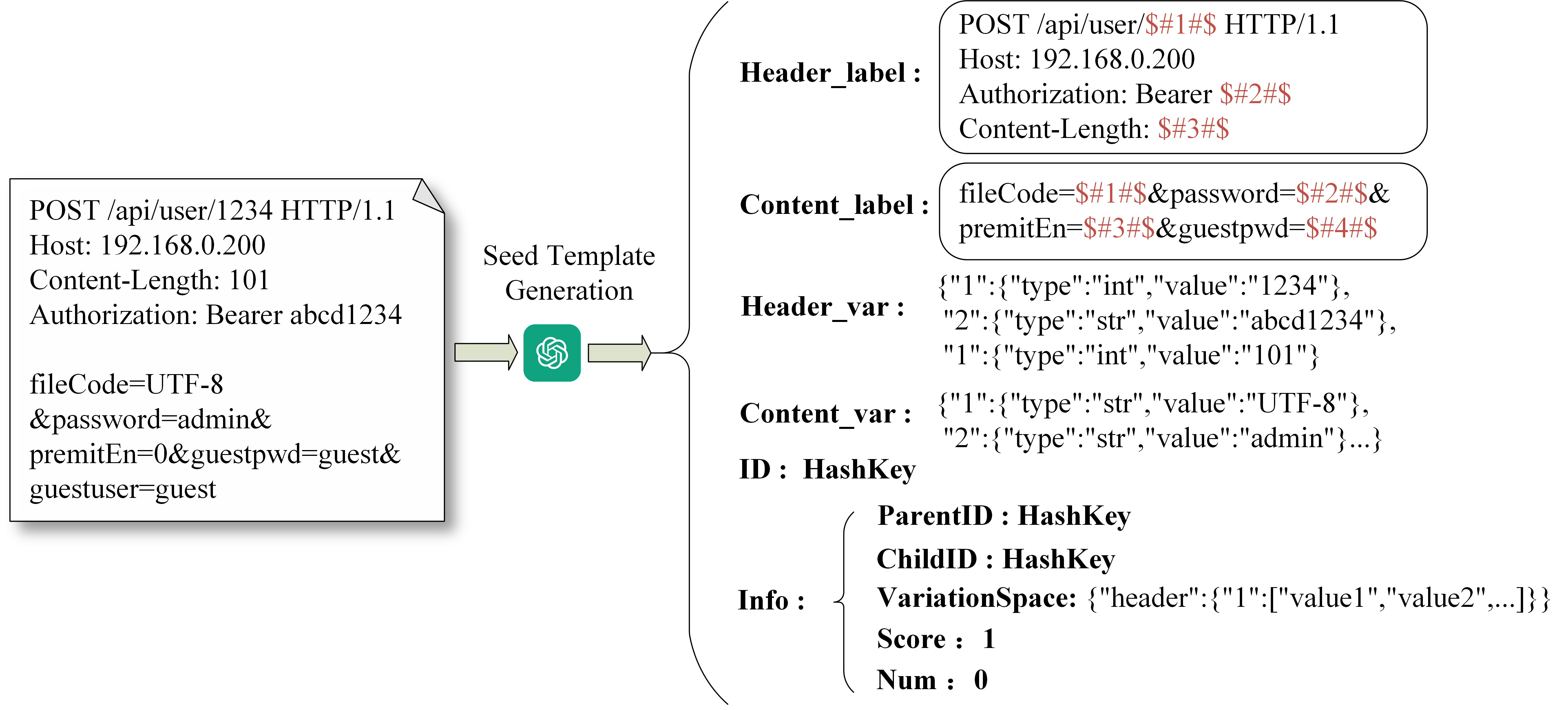}
\caption{Seed Template Struction. Showcasing HTTP protocol preservation, variable annotations, mutation guidance, and scheduling information.}\label{Seed Template Struction}
\end{figure*}

We generate seed templates using the packet information parsed by LLM.
As illustrated in Figure \ref{Seed Template Struction}, the seed templates preserve the structure of the HTTP protocol while extracting all variable parameter values. 
These templates document information related to seed mutations, enabling more effective guidance for seed scheduling and mutation processes. 
This structure mainly stores the following information.

\textbf{Header and Content Annotations.} Variables are identified and annotated within the header and content data, with the symbol $\$\#\#\$$ used to separate annotated data from the original HTTP data. 
This annotation method records the exact location of variables in the template, ensuring accurate replacement during the mutation process.

\textbf{Field Set.} To facilitate more targeted operations during mutation, the field set centrally manages all variable parameters, specifying the type of each field, such as integer (int), string (str), and special character encoding types (e.g., Base32, Base64, URL encoding).

\textbf{Mutation Space.} To fully leverage the code analysis capabilities of LLMs in fuzzing, we introduce a mutation space for each identified field, storing potential value assignments inferred by the LLM. 
This structure will be used in Section \ref{subsec4.2.3} to extend field values and in Section \ref{subsec4.3.2} to guide the fuzzer in performing context-aware intelligent mutations.

\textbf{Mutation Scoring.} Seeds that trigger more code branches and achieve higher scores are prioritized during scheduling by recording each round's seed score and periodically calculating scores based on response content and mutation space (Discuss in Section \ref{subsec4.4}).

\textbf{Invocation Count.} Each seed template's invocation count is tracked to ensure that templates with fewer invocations are given more testing opportunities.

\subsection{LLM-Guided Seed Template Enrichment}
\label{subsec4.2}
In this section, we focus on expanding the seed template library through two primary approaches: enriching data packets and extending field values.
We analyze key code segments and partial request contents using static analysis and carefully crafted prompts to generate additional effective seed templates and field values.
\subsubsection{Static Analysis Preprocessing}
\label{subsec4.2.1}

Before running ChatHTTPFuzz, static analysis preprocessing and asynchronous code analysis based on seed information can be performed during runtime. 
Specifically, the analysis focuses on the correspondence between the routing code and the backend code, which can be categorized into two modes:

The first mode is table-based routing. 
This approach uses a centralized routing table or configuration file to manage all routing information. 
Each URL path is explicitly mapped to a specific handler function or controller in such a structure. 
For example, in the following code, Tenda devices use the $add\_url$ function to register global routing information uniformly. 
The service code corresponding to the data packet can be identified by referencing the routing registration function.

\begin{lstlisting}[language=C]
add_url("GetWanStatistic",formGetWanStatistic);
add_url("SetRemoteWebCfg",formSetSafeWanWebMan);
...
\end{lstlisting}

The second mode is file-based routing. 
In this approach, the file system's structure typically determines the URL routing. 
Each URL path corresponds to a specific file, usually a CGI script or other executable file. 
For example, Dlink-DNS uses file-based routing, where the code is encapsulated in different files according to their functional roles, and each route corresponds to a CGI file.

\begin{lstlisting}[language=C]
/cgi-bin/account_mgr.cgi
/cgi-bin/addon_center.cgi
...
\end{lstlisting}

Once the routing method is determined, we use IDA Python \citep{noauthor_idapython_nodate} to extract the handler functions associated with the routing information. 
The decompiled code is then saved in the specified format within the folder configured by ChatHTTPFuzz, completing the static analysis preprocessing. 

\subsubsection{LLM-Guided Packet Generation}
\label{subsec4.2.2}
The initial method for capturing traffic involved intercepting packets through real-time page access. 
While this approach ensures real-time traffic acquisition, some packets associated with specific backend services may not be directly accessible via the frontend due to the complexity of the backend architecture. 
We utilize the routing information and code extracted in Section \ref{subsec4.2.1} to address these cases. 
Leveraging large language models (LLMs), we generate new request packets and employ protocol-aware techniques to parse the protocols, facilitating the creation of new seed templates.

\begin{figure}[!ht]
\centering
\includegraphics[width=1\linewidth]{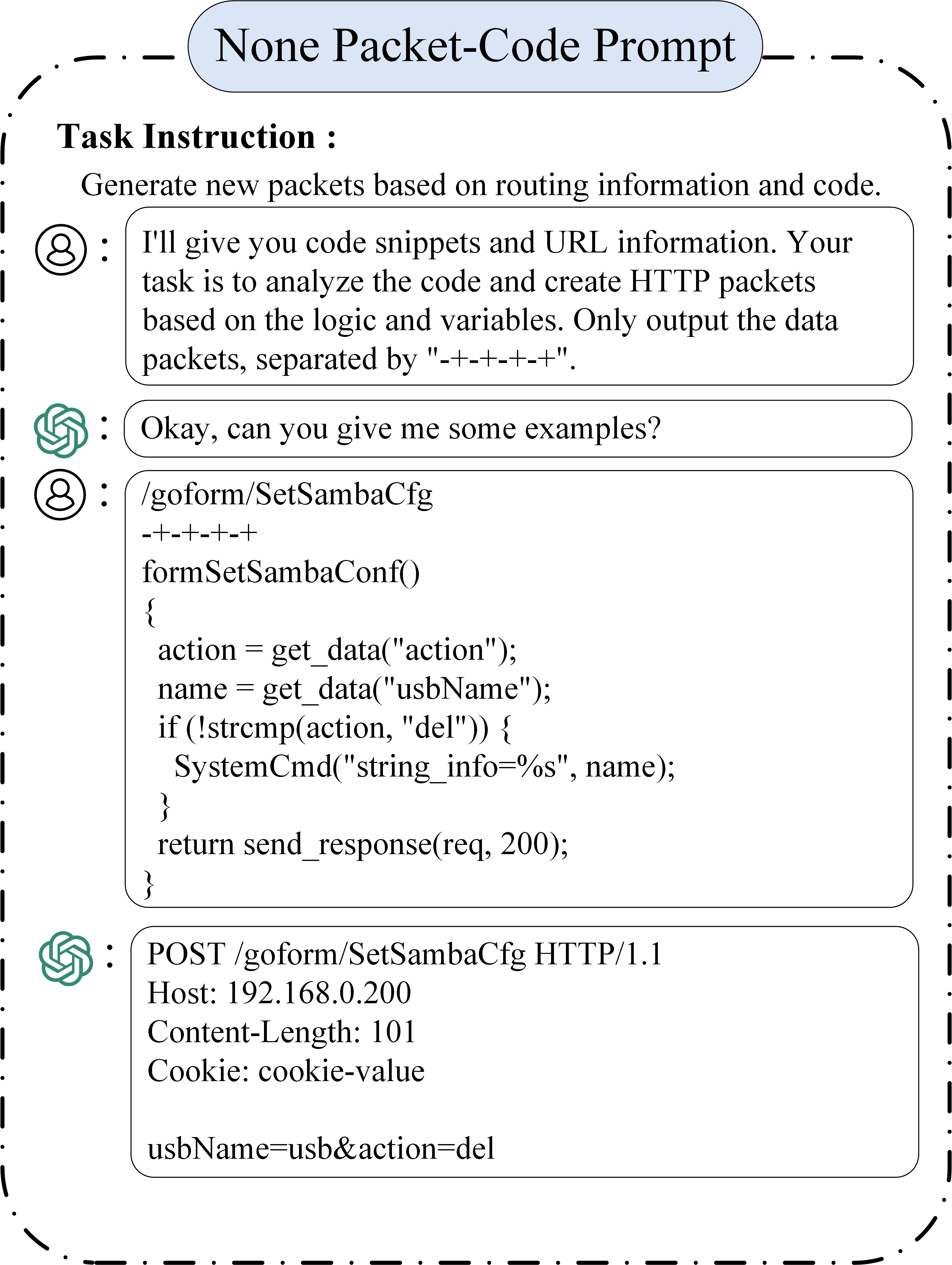}
\caption{Packet-Code Prompt.}\label{Packet-Code Prompt}
\end{figure}

Figure \ref{Packet-Code Prompt} describes that the guidance process utilizes a sequence of prompts to direct the LLM in parsing URLs and code snippets according to a specified format, using “-+-+-+-+” as a delimiter to ensure clarity and consistency. 
The backend logic typically depends on the values assigned to different fields, with these values triggering specific code blocks and requiring particular combinations to activate more complex logic paths. 
To comprehensively cover various execution paths, the prompt sequence provides precise instructions, enabling the LLM to generate multiple data packets that align with the code’s logic. 
This approach ensures that the generated packets correspond to the system's diverse logical branches, offering extensive support for further testing and analysis.

\subsubsection{LLM-Guided Field Expansion}
\label{subsec4.2.3}

For routes containing request packets, we leverage LLMs to ensure the diversity and accuracy of field values. 
This approach enables us to generate additional sets of fields that align with the code logic, referred to as the \enquote{Mutation Space} aimed at evaluating the potential for future seed mutations. 
The expansion of fields is achieved through the following two methods:

\begin{figure}[!ht]
\centering
\includegraphics[width=1\linewidth]{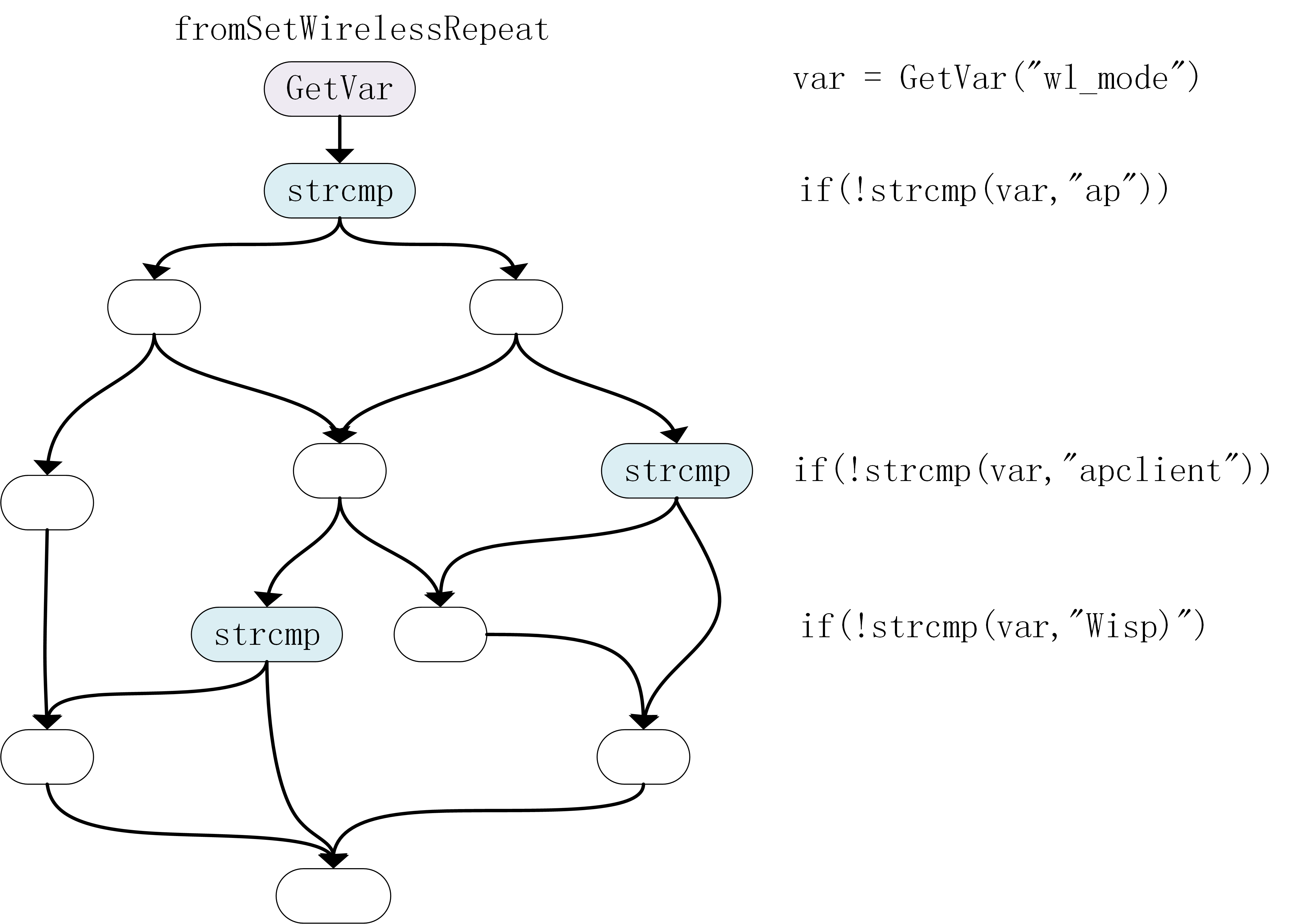}
\caption{IoT Code Block. Demonstrates the potential assignments of POST parameters to variables in the device code.}\label{IoT Code Block}
\end{figure}

The first involves identifying the assignable value sets for variables.
As shown in Figure \ref{IoT Code Block}, this backend code block for an IoT device uses the GetVar function to retrieve POST parameters from HTTP packets.
These parameters are compared using strcmp (or similar functions), with the code branching based on the results. 
Such comparison functions define the valid range or acceptable values for parameters.
By analyzing the code logic, large language models (LLMs) can identify these functions and extract the corresponding sets of valid strings. 
This process clarifies possible parameter values and generates logic-compliant packets to ensure test data triggers the intended execution paths.
We analyze four types of devices, evaluating the potential value sets of parameters within 30 backend code samples. 
According to the results presented in Table \ref{tab:field_enrich}, the LLM achieved an accuracy rate of 97.17\% in identifying these parameter value sets. 
This finding indicates that the LLM can accurately interpret the valid ranges of parameter values, providing effective guidance for generating test data that ensures variable assignments align with logical requirements and successfully trigger the intended execution paths.

\begin{table*}[!tb]
    \centering
    \caption{Field Value Recognition Rates for Different IoT Devices.}
    \label{tab:field_enrich}
    \small
    \begin{tabular}{lccc}
        \toprule
        \textbf{Device Name} & \textbf{Total Field Values Count} & \textbf{Total Recognized Count} & \textbf{Recognition Rate (\%)} \\
        \midrule
        \textbf{Dlink Dir-816} & 36 & 34 & 94.44\% \\
        \textbf{Tenda AC15} & 26 & 26 & 100\% \\
        \textbf{Dlink DNS320L} & 18 & 17 & 94.44\% \\
        \textbf{Zyxel} & 26 & 26 & 100\% \\
        \midrule
        \textbf{Total} & 106 & 103 & 97.17\% \\
        \bottomrule
    \end{tabular}
\end{table*}

The second focuses on determining specific format requirements. 
LLMs can be used to identify and extend the format requirements for fields within request packets. 
In addition to ensuring fields are assigned valid values, some parameters must adhere to strict format standards when generating packets. 
This format recognition and automatic generation capability enables ChatHTTPFuzz to create difficult field values for other state-of-the-art tools to construct, providing potential for discovering zero-day vulnerabilities.

In this study, we analyze the router code of several devices and find that format requirements are crucial for generating test packets.
For example, the following code snippet demonstrates the strict format constraints on certain fields:

\begin{lstlisting}[language=C]
char v22[80]; 
int v24;
int *v25;  
int *v26;
cgi = get_cgi((int)"services_array");
...
sscanf(cgi, "%d:%d:%d:%s", &v24, v25, v26, v22);
\end{lstlisting}

As demonstrated in the code above, variables v24, v25, and v26 are integers that must adhere to the \enquote{\%d} format requirement. At the same time, v22 is a character array with a maximum length of 80, requiring the \enquote{\%s} format specification. These parameters need to be separated by colons \enquote{:}. 
In addition, LLMs can recognize and extend special encoding formats such as Base64, date formats, and IP address formats, allowing for the generation of test packets that more closely follow real-world standards. 


\begin{algorithm}
\caption{Field Value Enrichment Algorithm}\label{algo2}
\begin{algorithmic}[1]
\Require $FuncAddr$: Address of fuzzing function, $URL$: URL name
\Ensure Enriched URL or seed templates
\Function{FieldEnrich}{$FuncAddr, URL$}
    \State $P_{values} \gets \texttt{StaticAnalysis}(FuncAddr)$ 
    \State $seed_{tps} \gets \texttt{GetSeedTemplate}(URL)$ 
    \If{$seed_{tps} \neq \varnothing$} 
        \ForAll{$seed_{tp} \in seed_{tps}$}
            \ForAll{$(key, values) \in P_{values}$} 
                \State $(type, index) \gets \texttt{Search}(key)$ 
                \State $seed_{tp}.\texttt{add}(type, index, values)$ 
            \EndFor
        \EndFor
    \EndIf
    \State \Return $seed_{tps}$
\EndFunction
\end{algorithmic}
\end{algorithm}

ChatHTTPFuzz utilizes the SeedTemplate structure to store the extended field sets obtained through the methods above, thereby providing a richer and more precise selection for seed mutations. 
For example, in the fromSetWirelessRepeat function of Tenda IoT devices, the field expansion process can be achieved through the Algorithm \ref{STSA}. 
First, the GetSeedTemplate function retrieves the seed template object associated with a specific URL.
Then, by iterating through the PotentialValues structure, the SeedTemplate's add function is called to add parameter values to the corresponding mutation list in the $VariationSpace$ variable within the SeedTemplate structure.

Using this field expansion algorithm, the $wl\_mode$ parameter in the fromSetWirelessRepeat function, after processing, supports multiple values including $ap$, $apclient$, and $wips$. 

\subsection{LLM-Guided Seed Mutation}
\label{subsec4.3}
Mutating arbitrary parts of the HTTP protocol often leads to the generation of numerous malformed packets. 
To enhance mutation efficiency, we integrate seed templates in the initial design phase. 
Seed templates identify and mark the mutable sections of HTTP packets. 
ChatHTTPFuzz first determines the type of the marked content and then applies targeted mutations to different fields. 
Additionally, it combines field values extracted from backend code using LLMs to perform further mutations (detailed in Section \ref{subsec4.4.2}). 
We also refer to common mutation methods proposed by Offutt \citep{offutt_experimental_2000}.
The marked fields undergo additional mutations using the Radamsa tool to introduce greater randomness and unpredictability, further diversifying the generated seed data.

\subsubsection{Type-based Mutation}
\label{subsec4.3.1}
For fields with numeric data types, the mutation operator modifies values to boundary or extreme cases (e.g., 0, -1, or extremely large numbers) to assess how the system behaves under extreme conditions.

For fields that use special encoding formats (such as Base32, Base64, or URL encoding), this study adopts a process of decoding, mutating, and then re-encoding. 
This approach ensures that the integrity and accuracy of the data are maintained during testing, regardless of the encoding format.

\subsubsection{Intelligent mutation combined with contextual content}
\label{subsec4.3.2}

ChatHTTPFuzz utilizes a combination of Large Language Models (LLMs) and the SeedTemplate structure to mutate field values intelligently to achieve more precise testing and improve code coverage. 
In this process, the LLM first analyzes the backend code associated with the HTTP packet to extract the actual usage and value sets for the fields Section \ref{subsec4.2.3}. 
ChatHTTPFuzz then performs targeted mutations based on the $VariationSpace$ variable, which contains the list of possible backend field values.

\subsubsection{Other variations}
\label{subsec4.3.3}

Handling Special Characters. During testing, special characters in common encoding formats (e.g., ASCII, Unicode, UTF-8, GBK) are replaced, along with null bytes (\textbackslash 0), newlines (\textbackslash n), carriage returns (\textbackslash r), and tabs (\textbackslash t). 
These special characters may cause unexpected parser behavior, potentially exposing hidden vulnerabilities.

Mutation Dictionary. To improve testing efficiency, a mutation dictionary can be set up in advance. 
This dictionary contains known exploit payloads that can be randomly selected during mutation testing. 
For example, payloads for command injection, buffer overflows, and integer overflows can be included to target specific vulnerabilities.

Radamsa-based Program Mutation. Radamsa is a powerful fuzzing tool that mutates input data in various ways to generate test cases. 
These test cases cover a wide range of boundary conditions and anomalous inputs.

\subsection{Seed Scheduling: Enhanced Thompson Sampling Method Based on Dual-Factor Gain}
\label{subsec4.4}

Seed scheduling is a crucial technique in fuzzing, aimed at optimizing testing outcomes by intelligently selecting and managing input seeds to improve efficiency and code coverage.
After generating many seed templates, efficient scheduling of these seeds is essential to balance  \enquote{exploration} and \enquote{exploitation}. 
This balance ensures that seeds are likely to trigger new branches and that untested seeds are more likely to be selected.

\enquote{exploration} refers to testing seed templates that have not been fully evaluated to gather more information and discover potentially effective templates. 
On the other hand, \enquote{exploitation} focuses on selecting the best-performing seed templates based on existing data to maximize immediate gains. 
This trade-off mirrors the core challenge addressed in multi-armed bandit problems within reinforcement learning \cite{slivkins_introduction_2024}, which aims to optimize the balance between exploration and exploitation.

\begin{figure*}[!th]
\centering
\includegraphics[width=0.9\linewidth]{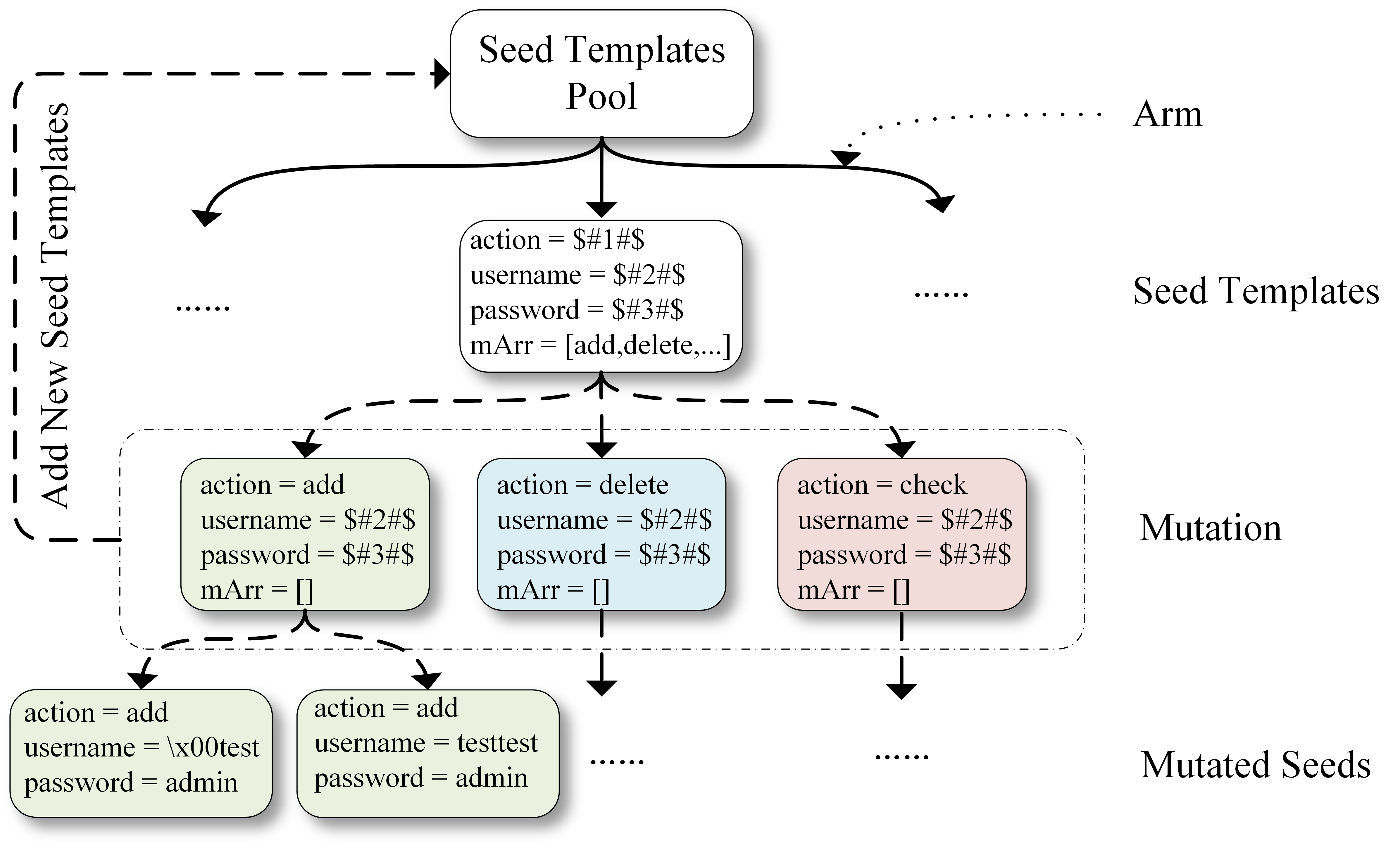}
\caption{Seed Template Relationship Structure. The diagram illustrates seed template selection and mutation process, with those generating new responses or triggering vulnerabilities added to the seed template pool.}\label{Seed Template Relationship Structure}
\end{figure*}

The structural relationship of seed templates is illustrated in Figure \ref{Seed Template Relationship Structure}, with the seed template repository positioned at the top level. 
During each usage, the system utilizes a template scheduling algorithm to select templates from the repository that require mutation. 
The mutation process is carried out based on the field types and their corresponding mutation spaces. 
Newly generated response templates are stored back into the seed template repository.
As the number of templates in the repository increases, we have developed a seed scheduling algorithm tailored to the structure of the seed templates. 
This algorithm aims to enhance the efficiency and logic of template selection, ensuring optimal utilization as the repository grows.

This study presents an innovative extension of the Thompson sampling algorithm by incorporating the field mutation space generated through the LLM-guided field extension technique outlined in Section \ref{subsec4.2.3}. 
Based on this approach, we propose a novel seed scheduling algorithm designed to optimize the utilization of seed templates. 
The algorithm achieves improved management and efficient scheduling of various seed templates through multidimensional evaluation and dynamic balancing strategies.

Our approach prioritizes three categories of seed templates with exceptional value to maximize their impact and ensure logical coherence and optimal system performance during the scheduling process.

$\bullet$
\textbf{Historically Successful Mutation Templates.} These templates have previously generated effective test cases, successfully triggering new program behaviors or uncovering potential vulnerabilities. Prioritizing these templates can significantly enhance testing efficiency and increase vulnerability detection rates.

$\bullet$
\textbf{Low-Frequency Invocation Templates.} This category includes templates that have been overlooked during multiple selection rounds for various reasons, as well as newly added templates to the template pool. Focusing on these templates helps maintain diversity and comprehensiveness in testing.

$\bullet$
\textbf{Rich Mutation Space Templates.} These templates contain multiple mutable fields or complex structures, offering broader exploration possibilities. Prioritizing such templates can expand testing coverage and depth.

For the first category, we utilize a seed selection algorithm based on Thompson sampling. 
This approach updates the Bayesian posterior probability based on historical performance, balancing exploring new and exploiting previously successful templates. 
For the second category, we use an exploration balance factor to target templates with lower invocation frequency. 
This ensures the system does not overlook these templates, especially newly added ones.
For the third category, we introduce a mutation potential factor, which predicts the future performance potential of templates, enhancing testing innovation and the capability to discover complex vulnerabilities.

\subsubsection{Seed selection algorithm based on Thompson sampling}
\label{subsec4.4.1}

To effectively utilize historical data, we employ the Thompson Sampling method. 
This method dynamically updates the Bayesian posterior distribution to balance the trade-off between exploration and exploitation, thereby improving the efficiency and accuracy of decision-making. 

Assume that the prior distribution of the \(N\) seed templates is denoted by \(\pi_i\). During the \(k\)-th round, the fuzz tester collects feedback information, represented as \( {Info}_k = (s_{0k}, s_{1k}) \), where \( s_{0k} \) indicates the number of attempts with a reward of 0, and \( s_{1k} \) denotes the number of attempts with a reward of 1 (where a reward signifies successfully triggering a new branch or discovering a vulnerability). 
Based on the performance of each template in the first \(k\) rounds, the posterior distribution is updated as \( \pi(\theta_i \mid {Info}_k)_{i \in (1, N)} \) for the \(k+1\)-th round, enabling the prediction of the performance of \( \theta_i \) in the subsequent round.

\subsubsection{Explore Balancing Factor}
\label{subsec4.4.2}

Maintaining a balance between exploration and exploitation during seed template selection is crucial for ensuring the effectiveness of fuzzing tests. 
To this end, we propose a novel Exploration Balance Factor (EBF) to optimize the template selection strategy. 
The core concept of this factor is to prioritize templates with low invocation frequency while preventing the testing process from being completely dominated by rarely used templates. 
We define the EBF as follows:

\begin{equation}
E_{factor_i} = \frac{log(total\_trials/N+1)}{log(\alpha_i+\beta_i+1)}.
\end{equation}

$total\_trials$ represents the total number of invocations for all templates, $N$ is the total number of templates, and $\alpha_i$ and $\beta_i$ denote the successful and unsuccessful invocations of template $i$, respectively. 
This design incorporates several key considerations: First, by placing the number of template invocations ($\alpha_i + \beta_i$) in the denominator, an inverse relationship is established between the $E_{factor_i}$ and the invocation frequency, thereby prioritizing the selection of less frequently used templates.
Second, using a logarithmic function effectively smooths extreme values, preventing templates with very low invocation counts from dominating the selection process excessively. 
Finally, the introduction of the $total\_trials/N$ term in the numerator allows the factor to adjust as the testing process progresses dynamically.
By incorporating this factor, our algorithm assigns a higher selection probability to templates with fewer invocations, thereby promoting the exploration of under-tested areas.

\subsubsection{Mutation Potential Factor}
\label{subsec4.4.3}

 The Mutation Potential Factor is a key metric in our algorithm for assessing the potential variability of seed templates. 
 This factor measures the potential mutation space of a template by comparing the logarithmic ratio of the current template's possible value set for each field to the maximum value set across all templates. 
 The calculation formula is as follows:

\begin{equation}
M_{factor_i} = \frac{\log(v_i + 1)}{\log\left(\sum_{j=1}^{N} v_j + 1\right)}.
\end{equation}

The numerator indicates the mutation potential of the current template, while the denominator represents the maximum potential among all templates.
$v_i$ Represents the size of the Variable Field Set for the current template $i$, $\sum_{j=1}^{N} v_j$ represents the total sum of the Variable Field Set across all templates. 

Using a logarithmic ratio prevents templates with many fields from dominating the selection process (e.g., one template may have hundreds of possible values while another has only a few).

The $M_{factor}$ indicates the potential value that a template may contribute in future tests. 
A higher $M_{factor}$ value suggests that, although the current reward for a template may not be the highest, it still has considerable room for improvement and potential value compared to other templates. 
Particularly in long-term fuzzing processes, the Mutation Potential Factor helps the algorithm avoid getting trapped in local optima and continues to explore templates that may lead to breakthrough discoveries.

\subsubsection{Integration}
\label{subsec4.4.4}

In this study, we propose a novel dual-factor gain enhanced Thompson Sampling method to optimize the seed template selection and scheduling process. 
By introducing the Mutation Potential Factor ($M_{factor}$) and the Exploration Balance Factor ($E_{factor}$), this method enhances the use of heuristic information while preserving the inherent advantages of traditional Thompson Sampling, resulting in a more efficient and balanced template selection strategy.

Our core innovation lies in the design of a dynamic gain mechanism, which adjusts and amplifies the traditional Thompson Sampling outcome ($\theta_i$) using $M_{factor_i}$ and $E_{factor_i}$, thereby expanding the dimensions considered in the decision-making process. 
Specifically, $M_{factor}$ evaluates the potential mutation capability of a template, enabling the algorithm to proactively consider the potential value of certain templates in future iterations. 
$E_{factor}$, on the other hand, is based on the historical usage of the template, ensuring an effective balance between exploration and exploitation during the decision-making process. 
The combined effect of these two factors allows the decision-making process to consider historical performance (as represented by $\theta_i$) and account for future potential and global balance. 
The specific integration approach is as follows:

\begin{equation}
Score_i = \theta_i \cdot (1+M_{factor_i}) \cdot (1+E_{factor_i}).
\end{equation}

We combine these factors with multiplication rather than addition in designing the gain mechanism. 
This choice creates a nonlinear gain mechanism that can more accurately capture the complex interactions between factors, especially in extreme cases. 
For instance, when the values of $M_{factor_i}$ or $E_{factor_i}$ are high, they amplify the impact of $S_i$, increasing the selection probability of certain templates. 
Conversely, when these factors are close to zero, they have minimal effect on the original Thompson Sampling results, ensuring the fundamental stability of the algorithm.

\begin{figure*}[!ht]
\centering
\includegraphics[width=0.9\linewidth]{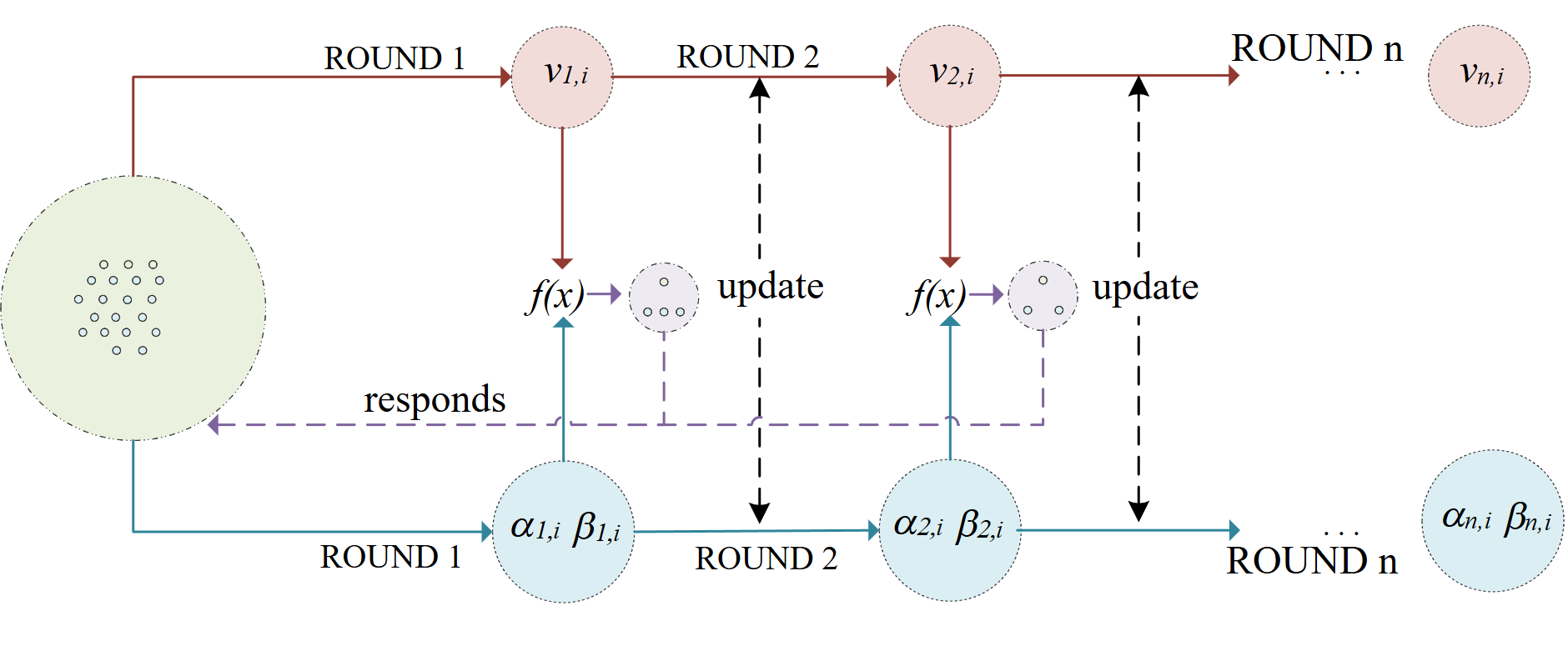}
\caption{Two-factor gain seed selection algorithm. 
The diagram illustrates the scoring process driven by two factors. 
The red section represents mutation potential, while the blue represents exploration balance. 
The system selects seed templates prioritized for mutation through iterative calculations in each round.}\label{Two-factor gain seed selection algorithm}
\end{figure*}

\begin{algorithm}
    \caption{Thompson Sampling Seed Selection Algorithm}
    \label{STSA}
    \begin{algorithmic}[1]
        \Require $T_s$: Seed Templates, a set of candidate arms
        \Ensure $S$: Top-selected arms based on Thompson Sampling

        \Function{SchedulingStrategy}{$T_s$}
            \If {$T_s = \emptyset$} 
                \State \Return $\emptyset$
            \EndIf

            \For{each $a_i \in T_s$}
                \State $(\alpha_i, \beta_i, v_i) \gets \Call{InfoExtract}{a_i}$
                \State $\theta_i \sim \text{Beta}(\alpha_i, \beta_i)$
                \State $E_i \gets \Call{ExploreBalancing}{\alpha_i, \beta_i}$
                \State $M_i \gets \Call{MutationPotential}{v_i}$
                \State $score_i \gets \Call{Score}{\alpha_i, \beta_i, v_i}$
            \EndFor

            \State $S \gets \text{Top}_{0.1N} \{score_i\}_{i=1}^N$
            \State \Return $S$
        \EndFunction
    \end{algorithmic}
\end{algorithm}

The implementation process is illustrated in Figure \ref{Two-factor gain seed selection algorithm} and Algorithm \ref{STSA}. 
In each iteration, key metrics for each seed template are first calculated, including the size of the field value set ($v$), the number of positive feedbacks ($\alpha$), and the number of negative feedbacks ($\beta$).
These metrics generate a composite $Score$ for each template using the designed scoring formula. 
The algorithm then selects templates with higher scores from the seed pool for further mutation, and these mutated templates are used for testing. 
During testing, feedback information such as triggered new responses and potential vulnerabilities discovered is updated in the seed pool to optimize decisions in the next iteration, forming a dynamic closed-loop optimization system.
\section{Implementation}
\label{sec:impl}
We develop a prototype of ChatHTTPFuzz, consisting of approximately 2000 lines of Python code, complemented by over 100 lines of Java code for the Burp Suite plugin. 
To improve processing efficiency and system scalability, we implement an asynchronous architecture in the design of this model.
This architecture allows system modules to execute their functions independently, optimizing overall performance and response time. 
High-speed message queues facilitate data transmission between modules, ensuring focused and efficient data processing. 
Additionally, the system architecture is divided into the following core modules to support efficient data processing and task execution:

\textbf{Traffic Capture and Analysis.} During the initial stage of traffic acquisition, a Burp Suite plugin is used to intercept and upload all traffic accessing IoT device web pages to a message queue, ensuring real-time data capture. 
This module analyzes traffic characteristics to determine whether new seed templates must be generated.

\textbf{LLM Protocol Awareness and Seed Template Generation.} LLM is employed to parse the HTTP protocol structure, identifying and labeling key variables in the protocol to generate seed templates with well-defined variable structures and types. 
These LLM-aware seed templates are crucial in the template compiler module, guiding seed scheduling and mutation more effectively.

\textbf{LLM Seed Template Expansion.} ChatHTTPFuzz performs static analysis on backend service code to extract routing information and request parameter names, which is the basis for expanding packets and field values for different routes. 
This approach follows the logical branches of the backend service code, generating more seed templates capable of triggering specific behaviors, thereby enhancing the breadth and depth of fuzzing.

\textbf{Template Mutator.} The mutator performs mutations based on the seed templates, combining random and rule-based approaches to mutate variables according to their types and contextual content. 
Mutation strategies are detailed in section 4.4.

\textbf{Feedback-Based Scheduling Strategy.} The strategy uses the response content of seeds to determine if new code paths have been triggered and employ reinforcement learning algorithms to balance exploration and exploitation. During exploration, the focus is on under-tested seed templates, while in the exploitation phase, templates with the best performance are prioritized. 
The strategy dynamically updates each seed template's mutation's success and failure counts. 
It quantifies their effectiveness using a Beta distribution score, optimizing information utilization and guiding subsequent scheduling decisions.

\textbf{Authentication Maintenance.} Since most IoT device code requires authentication for access, ChatHTTPFuzz continuously monitors session token validity and updates it as necessary to ensure that the session remains active, maximizing the testing coverage of backend service code.

\textbf{Anomaly Monitoring.} To detect various anomalies such as command injection and buffer overflow, ChatHTTPFuzz implements specialized detection strategies for each type. 
For command injection, each mutated packet embeds a unique seed template identifier into the command to be executed. 
The IoT device's detection code correlates the executed command with the seed template. 
For buffer overflow detection, each test connection captures the \enquote{Connection reset} exception, which often indicates a memory corruption vulnerability causing service disruption when the server disconnects.
\section{Experiment and Evaluation}
\label{sec6}
In this section, we evaluate the performance of ChatHTTPFuzz across various metrics through practical testing on IoT devices and conduct a comparative analysis with other tools.

\subsection{Experimental Setup}
\label{sec6.1}

Host and Device Configuration. ChatHTTPFuzz was set up on Ubuntu 22.04, with an Intel Core i9-12900KF 3.20 GHz CPU and 8GB RAM.

IoT Devices. We conduct tests on various mainstream brands, including Cisco, D-Link, TP-Link, Linksys, VIVOTEK, and Netgear, focusing on their routers, network storage devices, VPNs, and firewalls (The specific list of tested products is presented in Table \ref{tab:device_info}.)
The criteria for selecting these devices include the brand's market influence, the number of devices used, and the diversity and coverage of their network service functionalities. 
To ensure comprehensiveness and representativeness in our testing, we select eight enterprise-grade devices from well-known brands (such as Cisco, TP-Link, D-Link, and Netgear), primarily focusing on routers, firewalls, VPNs, and cameras related to network and security services. 


\begin{table}[!tb]
    \centering
    \caption{Device Information by Vendor.}
    \label{tab:device_info}
    \small
    \begin{tabular}{lccc}
        \toprule
        \textbf{Vendor} & \textbf{Device} & \textbf{Version} & \textbf{Type} \\
        \toprule
        Cisco & RV110W & V1.2.2.5-8 & VPN \\
        \multirow{5}{*}{D-Link} & Dir-816 & V1.10 & Router \\
                                 & DNS-320 & V2.05b08 & NAS \\
                                 & DNS-320L & V1.09b06 & NAS \\
                                 & DNS-327L & V1.10 & NAS \\
        \multirow{2}{*}{Tenda} & AC15 & V15.03.05.19 & Router \\
                                & AC10 & V16.03.10.13 & Router \\
        \multirow{2}{*}{TP-Link} & WR841N & V3.16.9 & Router \\
                                  & WR940N & v3.20.1 & Router \\
        Linksys & WRT54G & V4.21.5 & Router \\
        \multirow{2}{*}{Netgear} & WNAP320 & V2.0.3 & AP \\
                                  & R9000 & V1.0.4.26 & Router \\
        VIVOTEK & CC8160 & 0113b & Camera \\
        TotoLink & NR1800X & V9.1.0u.6279 & Router \\
        zyxel & NAS326 & V5.21(AAZF.18) & NAS \\
        \bottomrule
    \end{tabular}
\end{table}

Comparison of Tools. We conduct performance comparison experiments between ChatHTTPFuzz and other black-box fuzzing tools such as Boofuzz, Snipfuzz, and Mutiny. 
Each of these tools has its unique advantages, and the specific reasons for selection and the comparative experiments' details can be found in Section \ref{sec6.3}.

LLM Configuration. 
The LLM used in this study is based on the GPT-4o API. 
Compared to other large language models (LLMs), GPT-4o typically exhibits superior text generation performance, particularly in generation quality and contextual understanding. 
The template parameters within the GPT model significantly influence the creativity of the generated text. 
We typically set the temperature parameter to zero to ensure the accuracy of information extracted from events. 
This configuration helps minimize noise interference, enhancing the precision of information extraction.

Manual Configuration. 
To ensure that the sessions remain active, each type and version of the device must retrieve session validation packets and login packets. 
To efficiently utilize the LSTE functionality, it is necessary to perform reverse engineering on the backend program using IDA Python scripts to obtain the code and routing information, which will then be provided to the LSTE interface.

\subsection{ChatHTTPFuzzing}
\label{sec6.2}
From the 15 different IoT device models mentioned above, we randomly select 4 devices and prepare request data packets in advance for testing each device. 
To evaluate the impact of incorporating large language models (LLMs) for guiding seed template generation and expansion techniques (LSTE) and seed template scheduling strategies (STSA) within ChatHTTPFuzz on vulnerability discovery effectiveness and efficiency, we develop three comparative experimental models: ChatHTTPFuzz-NoLSTE, ChatHTTPFuzz-NoSTSA, and ChatHTTPFuzz-NoLSTE-NoSTSA. 
All these models retain the perception capabilities of LLMs, as testing cannot be conducted without the ability to generate seed templates.

\begin{figure}[t]
\centering
\includegraphics[width=1\linewidth]{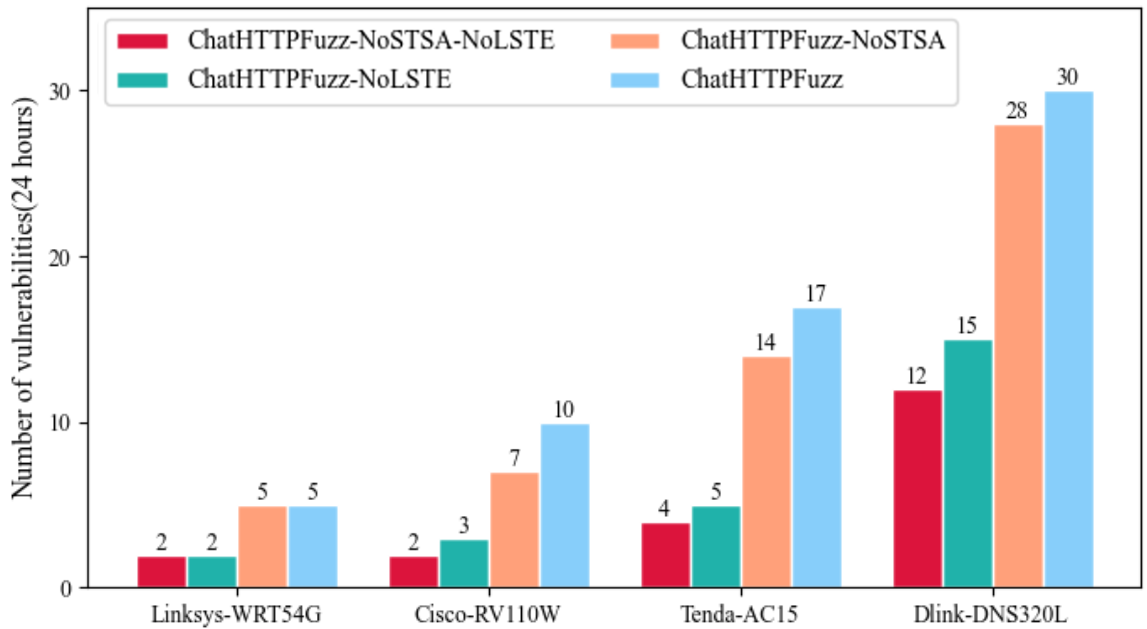}
\caption{Number of Vulnerabilities Detected by Different Versions of ChatHTTPFuzz.}\label{Number of Vulnerabilities Detected by Different Versions of ChatHTTPFuzz}
\end{figure}

Impact on Vulnerability Discovery. By leveraging LLMs to expand the seed template library, the ChatHTTPFuzz system enhances its understanding of backend code logic, enabling the generation of more logic-compliant seed templates and thereby discovering more vulnerabilities. 
As shown in Figure \ref{Number of Vulnerabilities Detected by Different Versions of ChatHTTPFuzz}, there is a significant difference in the total number of vulnerabilities discovered by ChatHTTPFuzz compared to ChatHTTPFuzz-NoLSTE across various IoT devices. 
Specifically, ChatHTTPFuzz identified an average of 2.38 times more vulnerabilities than ChatHTTPFuzz-NoLSTE across the four devices (57 vulnerabilities versus 24 vulnerabilities).

The seed template scheduling technique has a relatively minor impact on the vulnerability detection capability of the ChatHTTPFuzz system. 
As shown in Figure \ref{Number of Vulnerabilities Detected by Different Versions of ChatHTTPFuzz}, there is no significant difference in the number of vulnerabilities discovered between ChatHTTPFuzz-NoSTSA and ChatHTTPFuzz. 
Although the seed scheduling technique plays a crucial role in prioritizing the selection of seed templates, its influence on the number of vulnerabilities discovered diminishes over time and eventually becomes negligible.

\begin{figure*}[t]
\centering
\includegraphics[width=1\linewidth]{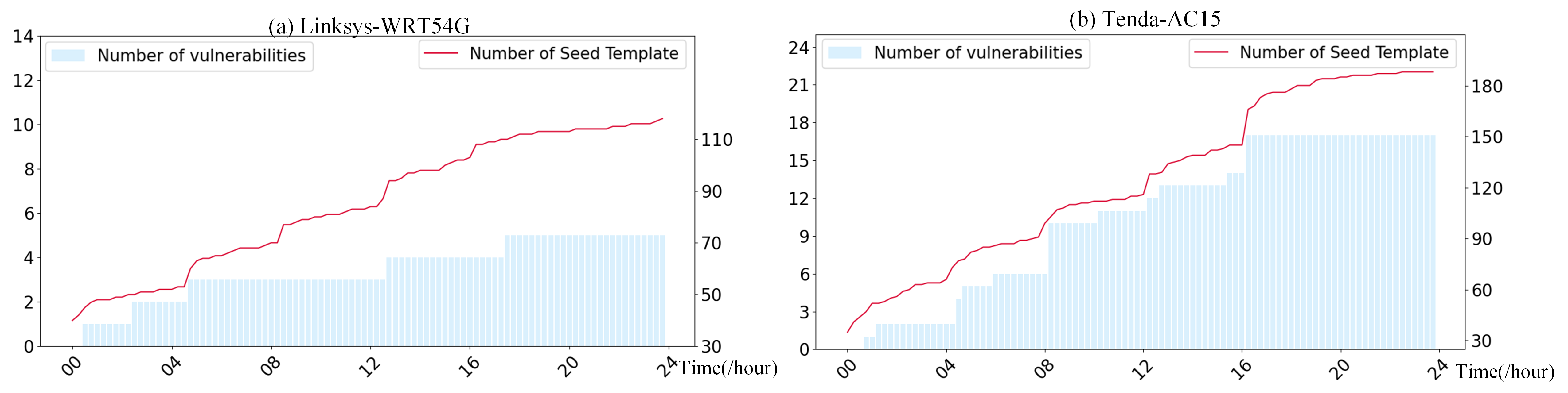}
\caption{Seed Template and Vulnerability Count Statistics Over a 24-Hour Period.}\label{Seed Template and Vulnerability Count Statistics Over a 24-Hour Period}
\end{figure*}

In practical testing, we observe that the number of seed templates significantly impacts the efficiency of vulnerability detection. 
To further validate this observation, we design an experimental setup focusing on two devices: the Tenda-AC15 and Linksys-WRT54G. Over a 24-hour testing period, we apply the LSTE technique every 4 hours to expand the existing seed template pool, 
while continuously monitoring the number of triggered vulnerabilities. 
The experimental results are illustrated in Figure \ref{Seed Template and Vulnerability Count Statistics Over a 24-Hour Period}.

The experimental results, shown in Figure \ref{Seed Template and Vulnerability Count Statistics Over a 24-Hour Period}, indicate that each application of the LSTE technique significantly increases the number of seed templates, accompanied by a rapid rise in the number of detected vulnerabilities.

\textbf{Impact on Vulnerability Detection Efficiency.}
By comparing the vulnerability detection time of ChatHTTPFuzz-NoSTSA and ChatHTTPFuzz over 24 hours, the results indicate that introducing the STSA technique significantly improved detection efficiency and, to some extent, increased the number of detected vulnerabilities. 
We initially used the LSTE technique to expand the seed template pool during the experiment. 
It was not employed during the subsequent fuzzing phase to prevent potential interference from the LSTE technique on the experimental results.

\begin{figure*}[!h]
\centering
\includegraphics[width=1\linewidth]{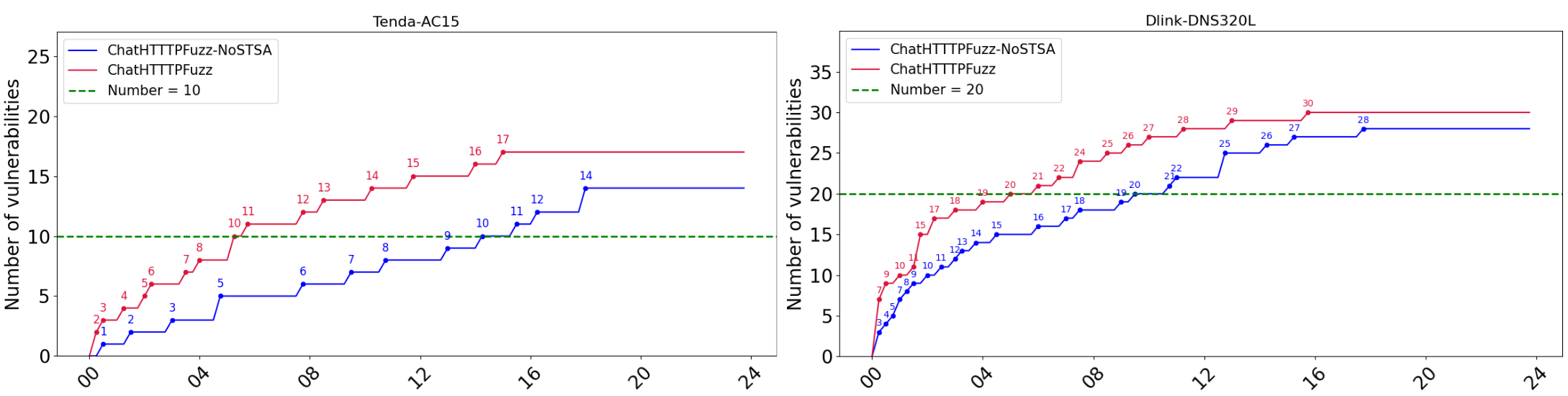}
\caption{Impact of STSA on Vulnerability Detection Efficiency.}\label{Impact of STSA on Vulnerability Detection Efficiency}
\end{figure*}

The experimental results, as illustrated in Figure  \ref{Impact of STSA on Vulnerability Detection Efficiency}, demonstrate that ChatHTTPFuzz reduces the time needed to discover the same number of vulnerabilities by half, compared to ChatHTTPFuzz-NoSTSA. 
This improvement is attributed to the STSA technique, which enhances the efficiency of seed template scheduling and selection.

\subsection{Comparison with Other Tools}
\label{sec6.3}
Various fuzzing tools are available for IoT devices, including recent research contributions such as mGPTFuzz, LABRADOR, and Snipuzz, as well as earlier benchmark tools like Boofuzz and Mutiny. 
Among these, mGPTFuzz is specifically designed for testing the Matter protocol in IoT devices, which is used for connected smart home systems \cite{zegeye_connected_2023}. 
Since LABRADOR has not been open-sourced, it was excluded from this study. 
The final selection of tools for the experiments is as follows.

\textbf{Boofuzz}, a branch and successor of the renowned Sulley fuzzing framework. 
It is specifically designed for protocol fuzzing, providing robust support and demonstrating its effectiveness in identifying various bugs in practical applications.
\textbf{Snipuzz}, a black-box fuzzing tool tailored for IoT devices. 
It optimizes message mutations by analyzing test responses, significantly improving error detection efficiency. 
Compared to traditional tools like Doona \cite{marcussen_wireghouldoona_2024} and IoTFuzzer \cite{chen_iotfuzzer_nodate}, Snipuzz exhibits superior performance in real-world testing, highlighting its advantages in security vulnerability detection.
\textbf{Mutiny}, an open-source fuzzer developed by Cisco researchers, utilizes a mutation-based approach for testing. 
It performs mutation fuzzing by replaying network traffic, greatly simplifying the testing process and enhancing convenience and efficiency. 
Additionally, the design of Mutiny makes it an ideal choice for conducting complex network protocol testing, effectively supporting a wide range of fuzzing applications.

To demonstrate the performance differences of LLMs in understanding protocol formats and selecting mutation points, we compare tools like Mutiny and Snipuzz for their mutation effectiveness. 
We design a set of HTTP packets with various parameter formats as the testing base. 
The experiments were divided into two groups: Group 1 used ChatHTTPFuzz to perform 1,000 mutation tests, assessing mutation point selection and effectiveness. 
Group 2 used tools like Mutiny and Snipuzz to run 1,000 tests on the same packets, enabling a comparative analysis.

\begin{figure}[!h]
\centering
\resizebox{\linewidth}{!}{
  \includegraphics{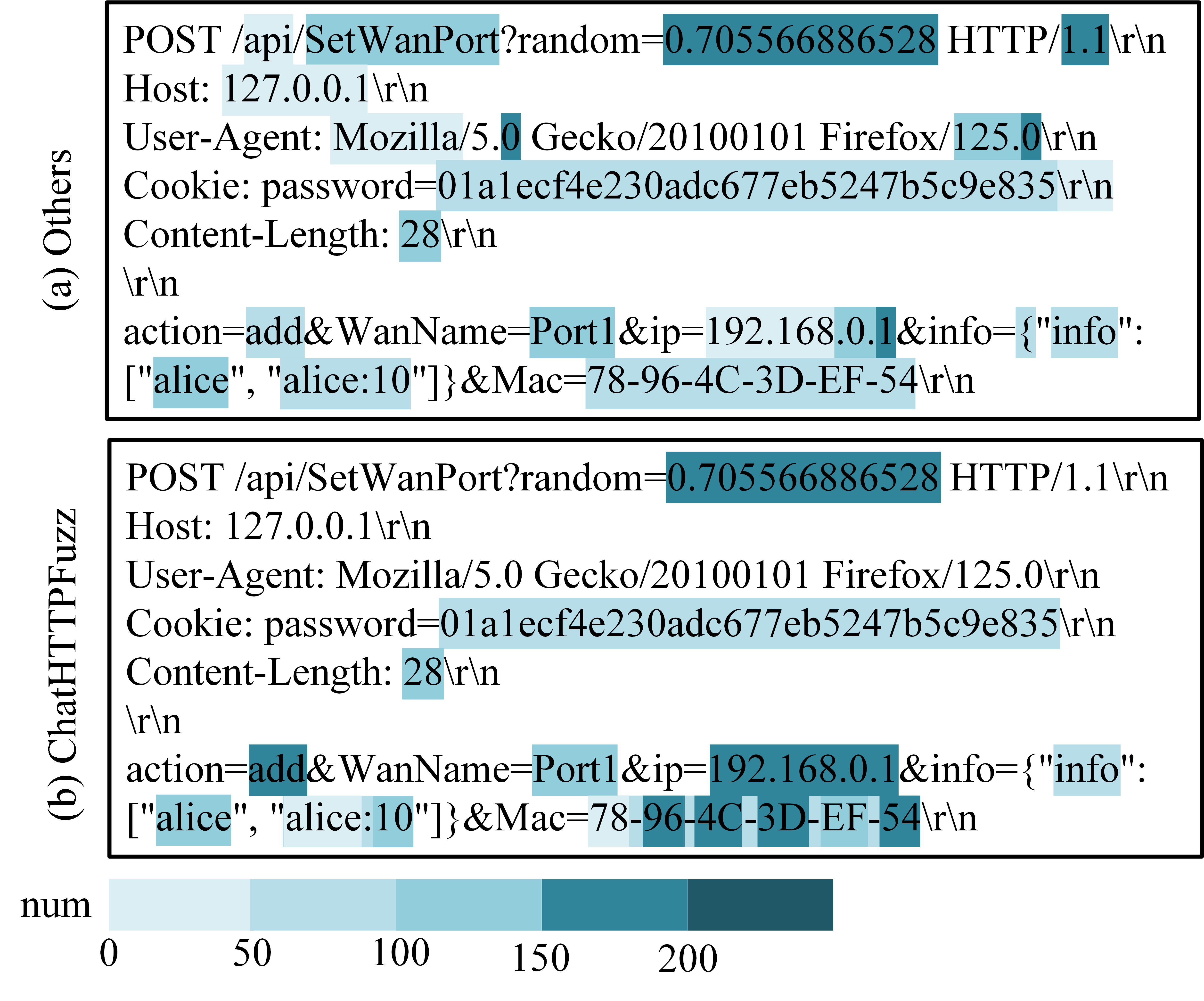}
}
\caption{Mutation Location Statistics. Each colored block in the diagram represents the mutation points selected by the fuzzing tool within the HTTP packet. The intensity of the color reflects the mutation frequency—darker shades indicate higher frequencies.}\label{filedlabel}
\end{figure}

The experimental results are shown in Figure \ref{filedlabel}, where each code block represents the mutation points selected by the fuzzing tools within the HTTP packets. 
The intensity of the block color reflects the mutation frequency—darker colors indicate higher mutation occurrences. 
The results demonstrate that ChatHTTPFuzz exhibits a clear advantage in protocol parsing and variable recognition. 
Specifically, compared to other tools, ChatHTTPFuzz more accurately avoids mutating fixed protocol elements (e.g., HTTP/1.1, $\backslash r \backslash n$
), which could compromise packet integrity and hinder parsing. 
Additionally, it outperforms tools like Mutiny and Snipuzz in handling complex variable structures by intelligently identifying variable boundaries and structures, enabling more targeted mutations and improving vulnerability detection.

\textbf{ChatHTTPFuzz VS Boofuzz.}
In comparing ChatHTTPFuzz and Boofuzz, we aim to evaluate LSTE technology's impact on other tools' vulnerability discovery capabilities. 
To this end, we apply LSTE technology as extensively as possible in our experiments with Boofuzz. 
LSTE enhances the quality of seed generation by leveraging the capabilities of large language models (LLMs) to produce more seeds that conform to business logic. 
The traditional Boofuzz tool heavily relies on manual annotation of fields that require mutation and cannot schedule seeds based on feedback. 
In testing five devices and seven firmware models, it detected only six publicly disclosed CVE vulnerabilities. 
However, Boofuzz augmented with LSTE technology (i.e., Boofuzz-LSTE) demonstrated significant advantages in vulnerability discovery; using tagged seeds, the number of vulnerabilities discovered was more than four times that of traditional Boofuzz. 
Nevertheless, due to Boofuzz's lack of mutation methods for command injection, it still failed to discover the command injection vulnerability in the Dlink DNS320L.


\begin{table}[!tb]
    \centering
    \caption{Comparison of Fuzzing Tools for Different IoT Devices. Fuzzer1: Boofuzz, Fuzzer2: Boofuzz-LSTE, Fuzzer3: ChatHTTPFuzz. }
    \label{tab:fuzzing_tools_comparison}
    \begin{tabular}{lcccc}
        \toprule
        \textbf{Vendor} & \textbf{Device} & \textbf{Fuzzer1} & \textbf{Fuzzer2} & \textbf{Fuzzer3} \\
        \toprule
        Dlink  & DNS320L & 0 & 0 & 32 \\
               & Dir816  & 3 & 8 & 13 \\
        Tenda  & AC15    & 1 & 9 & 17 \\
        TP-Link & WR841N & 1 & 1 & 1 \\
                & WR940N & 0 & 1 & 1 \\
        Linksys & WRT54G & 1 & 2 & 7 \\
        Cisco   & RV100W & 0 & 6 & 10 \\
        \toprule
        \textbf{Total} & & \textbf{6} & \textbf{25} & \textbf{81} \\
        \bottomrule
    \end{tabular}
    
\end{table}

\textbf{ChatHTTPFuzz VS Snipuzz.} 
Unlike SNIPUZZ, which primarily infers message fragments and refines mutation strategies based on response discrepancies, ChatHTTPFuzz leverages response variations and code context information to guide the mutation process and seed scheduling more intelligently. 
As a black-box fuzzing tool, SNIPUZZ lacks a deep understanding of the code, and its method of attempting to determine strings byte-by-byte during mutation often results in a large amount of redundant data. 
Furthermore, SNIPUZZ fails to adequately consider multiple types of vulnerabilities during the mutation process, limiting its vulnerability discovery capability. 
Table \ref{tab:vulnerability_discovery_time} illustrates the differences in vulnerability discovery capabilities between ChatHTTPFuzz and SNIPUZZ.

\textbf{ChatHTTPFuzz VS Mutiny.} 
Mutiny primarily relies on the Radamsa mutator. 
However, as a general-purpose mutator, Radamsa does not fully consider the context and structural information of the input data, which can result in the generation of test cases that lack specificity and ultimately reduce the efficiency of fuzzing.
Moreover, Radamsa’s random mutation strategy often generates a large amount of invalid input, which not only increases testing time and costs but also hinders the creation of high-quality test cases, thereby limiting its contribution to the discovery of new vulnerabilities. 
As shown in Table \ref{tab:fuzzing_tools_comparison}, there is a significant gap in vulnerability discovery capabilities between Mutiny and ChatHTTPFuzz within the same testing timeframe.

\begin{table*}[!tb]
    \centering
    \caption{Vulnerability Discovery Time Comparison for Different IoT Devices.}
    \label{tab:vulnerability_discovery_time}
    \small
    \begin{tabular}{lcccccc}
        \toprule
        \textbf{CVE ID} & \textbf{Type} & \textbf{Device} & \textbf{ChatHTTPFuzz} & \textbf{Boofuzz} & \textbf{Snipuzz} & \textbf{Mutiny} \\
        \midrule
        CVE-2020-8423 & Buffer Overflow & TP-Link & \rr & \rr & \ww & \rr \\
        CVE-2020-3331 & Buffer Overflow & Cisco RV110W & \rr & \ww & \ww & \rr \\
        CVE-2024-7439 & Buffer Overflow & Vivotek CC8160 & \rr & \ww & \ww & \ww \\
        CVE-2024-41281 & Buffer Overflow & Linksys WRT54G & \rr & \rr & \ww & \ww \\
        CVE-2024-2855 & Buffer Overflow & Tenda AC15 & \rr & \ww & \ww & \ww \\
        CVE-2024-2902 & Buffer Overflow & Tenda AC15 & \rr & \rr & \ww & \rr \\
        CVE-2019-20760 & Command Injection & NetGear R9000 & \rr & \ww & \ww & \ww \\
        CVE-2024-7922 & Command Injection & DLink-DNS 320L & \rr & \ww & \ww & \ww \\
        CVE-2024-8130 & Command Injection & DLink-DNS 320L & \rr & \ww & \ww & \ww \\
        CVE-2024-7715 & Command Injection & DLink-DNS 320L & \rr & \ww & \ww & \ww \\
        CVE-2022-41518 & Command Injection & ToTolink & \rr & \ww & \ww & \ww \\
        \bottomrule
    \end{tabular}
\end{table*}

Why can Chathttpfuzz only discover some vulnerabilities?

\textbf{(1) Detection of Multiple Types of Vulnerabilities.}
    ChatHTTPFuzz is configured with various vulnerability detection modules, including buffer overflow, command injection, file reading, and a mutation mechanism for Base64 encoding. 
    It successfully identified the CVE-2019-20760 vulnerability during the testing process, which other tools had not detected. 
    Additionally, in tests conducted on D-Link DNS devices, ChatHTTPFuzz discovered 30 similar vulnerabilities, including CVE-2024-7922, CVE-2024-8130, and CVE-2024-8131, which were also undetected by other tools.

\textbf{(2) Vulnerabilities Requiring Contextual Information.} 
    Certain vulnerabilities can only be exploited under specific contextual conditions.
    For example, the CVE-2024-2855 vulnerability in Tenda AC15 devices is only triggered when the timeType value is manually set. 
    However, the initial value of timeType is sync, necessitating an in-depth analysis of the code logic to find situations where manual is assigned. 
    Similarly, the CVE-2024-7715 vulnerability in D-Link DNS-320L devices is only triggered when the type parameter is set to photos. 
    These vulnerabilities are challenging to trigger without the support of advanced LSTE techniques.

\textbf{(3) Vulnerabilities Requiring Special Mutation Mechanisms.}
    To address the challenges posed by various encoding formats in mutation analysis, ChatHTTPFuzz incorporates parsing and wrapping mechanisms for common encoding formats such as hex, Base64, and URL encoding. 
    This capability allows it to recognize and perform fuzzing on content within these encoded formats. 
    ChatHTTPFuzz has successfully identified multiple vulnerabilities using this method, including CVE-2019-20760, CVE-2024-7439, and CVE-2024-41281.

\subsection{Vulnerability Discovery}
\label{sec6.4}

\begin{table*}[!tb]
    \centering
    \caption{Vulnerability Discovery Statistics for Different IoT Devices.}
    \small
    \label{tab:vulnerability_discovery_statistics}
    \begin{tabular}{lcccccc}
        \toprule
        \textbf{Vendor} & \textbf{Device} & \textbf{Version} & \textbf{Vul} & \textbf{Undisc. Vul} & \textbf{CVE} \\
        \midrule
        Cisco   & RV100W   & V1.2.2.5-8 & 13 & 11 & - \\
        \midrule
        TP-Link & WR841N   & V3.16.9 & 2 & - & - \\
                & WR940N   & V3.20.1 & 1 & - & - \\
        \midrule
        \multirow{4}{*}{D-Link} & Dir-816  & V1.10 & 12 & 9 & \multirow{4}{*}{21} \\
                & DNS-320  & V2.05b08 & 31 & 30 &  \\
                & DNS-320L & V1.09b06 & - & - &  \\
                & DNS-327L & V1.10 & - & - &  \\
        \midrule
        Tenda   & AC15     & V15.03.05.19\_multi & 17 & 2 & - \\
                & AC10     & V16.03.10.13 & 4 & - & - \\
        \midrule
        Linksys & WRT54G   & v4.21.5 & 8 & 7 & 2 \\
        \midrule
        Netgear & R9000    & V1.0.4.26 & 1 & - & - \\
        \midrule
        VIVOTEK & CC8160   & 0113b & 1 & - & - \\
        \midrule
        TotoLink& NR1800X  & V9.1.0u.6279\_B20210910 & 2 & - & - \\
        \midrule
        zyxel   & NAS326   & V5.21(AAZF.18)C0 & 11 & 9 & - \\
        \midrule
        \textbf{Total} & & & \textbf{103} & \textbf{68} & \textbf{23} \\
        \bottomrule
    \end{tabular}
\end{table*}

As shown in Table \ref{tab:vulnerability_discovery_statistics}, ChatHTTPFuzz successfully identifies 92 security vulnerabilities across eight different models of IoT devices, of which 59 were previously undisclosed. 
These vulnerabilities have been reported to the respective vendors and have been confirmed. Currently, 22 of these vulnerabilities have been assigned official CVE identifiers.

The vulnerabilities identified by ChatHTTPFuzz mainly fall into Command Injection (CI) and Memory Corruption. 
Command Injection is a high-risk logical vulnerability that can lead to Remote Code Execution (RCE) through specific attack techniques. 
It is generally more stable compared to memory corruption vulnerabilities. 
ChatHTTPFuzz discovers 31 command injection vulnerabilities in the tested IoT devices, 17 of which, including CVE-2024-7828, have been rated as high severity according to the CVSS V3 standard.

Due to the limited computational capacity and the simplified stack data handling typically found in IoT devices, memory corruption has become one of these devices' most common vulnerability types. 
We have identified 10 stack overflow vulnerabilities in Cisco products that could lead to Denial of Service (DoS) or remote code execution. 
However, since these products are at their end-of-life (EOL) stage, CVE identifiers could not be assigned. 
Additionally, in Linksys products, we found 7 memory corruption vulnerabilities, including CVE-2024-41281, which may affect the stability and control of the devices. 
Overall, ChatHTTPFuzz identifies 32 memory corruption vulnerabilities across six different IoT devices.

In summary, among the 22 CVE-confirmed vulnerabilities, 4.54\% are of medium severity, 13.64\% are of high severity, and 81.82\% are classified as critical severity. 
The results are shown in Table \ref{tab:cve_vulnerabilities}.

\begin{table}[!tbh]
    \centering
    \caption{CVE Vulnerabilities and Their CVSS v3 Scores.}
    \label{tab:cve_vulnerabilities}
    \begin{tabular}{lccc}
        \toprule
        \textbf{Vendor} & \textbf{Device} & \textbf{CVE ID} & \textbf{CVSS v3} \\
        \midrule
        Linksys & WRT54G & CVE-2024-41281 & 8.8 / High \\
        \midrule
        \multirow{21}{*}{D-Link} & \multirow{21}{*}{DNS-320L} & CVE-2024-7828 & 9.8 / Critical \\
                                 & & CVE-2024-7829 & 9.8 / Critical \\
                                 & & CVE-2024-7830 & 9.8 / Critical \\
                                 & & CVE-2024-7831 & 9.8 / Critical \\
                                 & & CVE-2024-7922 & 9.8 / Critical \\
                                 & & CVE-2024-8127 & 9.8 / Critical \\
                                 & & CVE-2024-8128 & 9.8 / Critical \\
                                 & & CVE-2024-8129 & 9.8 / Critical \\
                                 & & CVE-2024-8130 & 9.8 / Critical \\
                                 & & CVE-2024-8131 & 9.8 / Critical \\
                                 & & CVE-2024-8132 & 9.8 / Critical \\
                                 & & CVE-2024-8133 & 9.8 / Critical \\
                                 & & CVE-2024-8134 & 9.8 / Critical \\
                                 & & CVE-2024-8210 & 9.8 / Critical \\
                                 & & CVE-2024-8211 & 9.8 / Critical \\
                                 & & CVE-2024-8212 & 9.8 / Critical \\
                                 & & CVE-2024-8213 & 9.8 / Critical \\
                                 & & CVE-2024-8214 & 9.8 / Critical \\
                                 & & CVE-2024-7849 & 8.8 / High \\
                                 & & CVE-2024-7832 & 8.8 / High \\
                                 & & CVE-2024-7715 & 6.3 / Medium \\
        \bottomrule
    \end{tabular}
\end{table}
\section{Related Work}
\label{sec:related}

\subsection{Black-Box Fuzzing}
\label{sec7.1}
Currently, black-box fuzzing is one of the main methods for testing the security of IoT devices. 
However, existing methods have various limitations. 
For instance, BooFuzz relies on empirically generated protocol packet templates, leading to high labor costs. 
Mutiny performs mutation-based fuzzing by replaying network traffic, which reduces the need for human intervention but also loses the ability to recognize protocol fields. 
SNIPUZZ infers the field format of request data by analyzing response content, which increases mutation efficiency but decreases the precision of data understanding, resulting in a large number of invalid seeds.

Overall, the limitation of black-box fuzzing in the Internet of Things (IoT) lies in the lack of protocol format parsing and in-depth understanding of code behavior, which makes the mutation and selection of seeds less targeted.

\subsection{Grey-Box Fuzzing}
\label{sec7.2}

In grey-box fuzzing, code coverage is a key factor that guides seed selection and mutation. 
However, IoT devices rely on multiple hardware components to function properly, complicating the acquisition of code coverage. 
As a result, many advanced research directions have shifted towards using emulation and hardware-supported techniques to obtain coverage information internally.

TriforceAFL employs QEMU’s \citep{noauthor_qemuqemu_2024} full-system emulation technology to boot and run firmware while integrating the AFL algorithm to collect and feedback code coverage. 
However, full-system emulation suffers from significant performance bottlenecks, severely impacting fuzzing efficiency. 
To address this, FIRM-AFL inherits the full-system emulation concept from TriforceAFL and implements a switch between system emulation and user-mode emulation by integrating the Firmadyne \citep{tufano_towards_2021} IoT emulator with QEMU user mode. 
This hybrid emulation strategy effectively improves operational efficiency, making FIRM-AFL a high-throughput coverage-guided fuzzing (CGF) tool. 

ChatHTTPFuzz employs a non-intrusive approach utilizing large language models (LLMs) to gain deep insights into protocol structures and backend code. 
This understanding guides the generation and mutation of seeds, enabling comprehensive testing of web services across all IoT devices.

\subsection{Other Work}
\label{sec7.3}

With the rise of large language models (LLMs), fuzzing is undergoing a new wave of transformation. 
LLMs demonstrate impressive advantages in code parsing, protocol analysis, and text generation, with increasing applications in IoT and network protocols, gradually integrating with traditional fuzzing techniques. 
ChatAFL \citep{meng_large_2024} combines LLMs with traditional AFL (American Fuzzy Lop) fuzzing tools, utilizing LLMs to generate more diverse input data, thus enhancing coverage and vulnerability discovery efficiency. 
mGPTFuzz \citep{ma_one_2024} is based on the GPT model and uses natural language processing capabilities to understand and generate test data, enabling the automatic generation of test cases for complex protocols and significantly reducing the need for manual intervention regarding protocol structures. 
This tool also adjusts input generation strategies based on the feedback from test results, significantly improving testing efficiency and accuracy. 
LLMIF \citep{wang_llmif_2024} uses large language models to identify and understand key fields in network protocols, generating targeted fuzzing inputs, and has designed an efficient fuzzing model for the Zigbee \citep{ramya_study_2011} protocol.

ChatHTTPFuzz is a specialized tool designed for fuzzing the HTTP protocol in the context of the Internet of Things (IoT). 
Similar to advanced techniques, it leverages the powerful capabilities of large language models (LLMs) to enhance protocol parsing, seed generation, and mutation strategies. 
By comprehensively understanding protocol intricacies, ChatHTTPFuzz generates more precise test data, improving the efficiency of vulnerability detection, especially in complex HTTP protocol environments.
\section{Conclusions}
\label{sec:conclu}
In this study, we present ChatHTTPFuzz, an IoT device HTTP protocol fuzzing framework based on Large Language Models (LLM). 
It leverages LLM to annotate data fields, construct structured seed templates, and identify suitable positions for mutation. 
Subsequently, it analyzes service code logic to generate matching data packets, enhancing the quality and quantity of seeds and expanding the mutation space of templates and field values. 
Finally, a dual-factor gain model is introduced, integrated with an enhanced Thompson sampling algorithm, to enable intelligent scheduling of seed templates, significantly improving fuzzing efficiency.

We test 14 device models from 9 leading vendors in the market. 
The results demonstrate that ChatHTTPFuzz significantly outperforms existing state-of-the-art tools, such as BooFuzz, Snipuzz, and Mutiny, in terms of vulnerability discovery capability. 
ChatHTTPFuzz successfully identifies 68 zero-day vulnerabilities, including command injection and memory corruption, across 6 different device models. 
Among these vulnerabilities, 23 of these discovered vulnerabilities have been officially assigned CVE identifiers.

\section*{Acknowledgments}
This work is supported by the NSFC through grants 62322202 and 62432006, Local Science and Technology Development Fund of Hebei Province Guided by the Central Government of China through grant 246Z0102G, Hebei Natural Science Foundation through grant F2024210008, and Guangdong Basic and Applied Basic Research Foundation through grant 2023B1515120020.

\bibliographystyle{natbib}

\bibliography{ref}

\end{document}